\documentclass[preprint]{aastex}

\newcommand{\rs}{r_{_{\rm S}}}

\newcommand{\OmegaK}{\Omega_{\rm K}}

\newcommand{\gapprox}{\lower.4ex\hbox{$\;\buildrel
>\over{\scriptstyle\sim}\;$}}
\newcommand{\lapprox}{\lower.4ex\hbox{$\;\buildrel
<\over{\scriptstyle\sim}\;$}}
\newcommand{\begeq}{\begin{equation}}
\newcommand{\fineq}{\end{equation}}
\newcommand{\begfig}{\begin{figure}}
\newcommand{\finfig}{\end{figure}}
\newcommand{\begeqarray}{\begin{eqnarray}}
\newcommand{\fineqarray}{\end{eqnarray}}
\def\N0{\dot N_0}

\def\Lesc{L_{\rm esc}}
\newcommand{\Eesc}{E_{\rm esc}}

\newcommand{\green}{f_{_{\rm G}}}
\newcommand{\Ljet}{L_{\rm jet}}

\newcommand{\SgrA}{Sgr~A$^*$}
\newcommand{\speed}{u}
\def\pseudophi{\Phi}

\def\ellprime0{\ell'_0}

\slugcomment{accepted by \apj}

\shorttitle{Particle Acceleration in Shocked Disks}

\shortauthors{Le \& Becker}

\begin{document}

\title{Particle Acceleration in Advection-Dominated Accretion Disks
\break with Shocks: Green's Function Energy Distribution}

\author{Truong Le\altaffilmark{1}}

\affil{E. O. Hulburt Center for Space Research, Naval Research
Laboratory, \break Washington, DC 20375, USA}

\and

\author{Peter A. Becker\altaffilmark{2}$^,$\altaffilmark{3}}

\affil{Center for Earth Observing and Space Research, George Mason
University, \break Fairfax, VA 22030-4444, USA}

\altaffiltext{1}{truong.le@nrl.navy.mil}
\altaffiltext{2}{pbecker@gmu.edu}
\altaffiltext{3}{also Department of Physics and Astronomy,
George Mason University, Fairfax, VA 22030-4444, USA}

\begin{abstract}

The distribution function describing the acceleration of relativistic
particles in an advection-dominated accretion disk is analyzed using a
transport formalism that includes first-order Fermi acceleration,
advection, spatial diffusion, and the escape of particles through the
upper and lower surfaces of the disk. When a centrifugally-supported
shock is present in the disk, the concentrated particle acceleration
occurring in the vicinity of the shock channels a significant fraction
of the binding energy of the accreting gas into a population of
relativistic particles. These high-energy particles diffuse vertically
through the disk and escape, carrying away both energy and entropy and
allowing the remaining gas to accrete. The dynamical structure of the
disk/shock system is computed self-consistently using a model previously
developed by the authors that successfully accounts for the production
of the observed relativistic outflows (jets) in M87 and \SgrA. This
ensures that the rate at which energy is carried away from the disk by
the escaping relativistic particles is equal to the drop in the radial
energy flux at the shock location, as required for energy conservation.
We investigate the influence of advection, diffusion, and acceleration
on the particle distribution by computing the nonthermal Green's
function, which displays a relatively flat power-law tail at high
energies. We also obtain the energy distribution for the particles
escaping from the disk, and we conclude by discussing the spectrum of
the observable secondary radiation produced by the escaping particles.

\end{abstract}



\section{INTRODUCTION}

Accretion flows onto supermassive black holes with mass $M \gapprox
10^8\,M_\odot$ are believed to power the high-energy emission observed
from quasars and active galactic nuclei (AGNs), as discussed by
Lynden-Bell (1969; see also Novikov \& Thorne 1973; Rees 1984; and Ford
et al. 1994). When the accretion rate is relatively low compared with
the Eddington value, the flow is radiatively inefficient, and the gas
temperature approaches the virial value. In this case the disk has an
advection-dominated structure, and most of the binding energy is
swallowed by the black hole (e.g., Narayan, Kato, \& Honma 1997;
Blandford \& Begelman 1999; Becker, Subramanian, \& Kazanas 2001; Becker
\& Le 2003). Despite the inefficiency of X-ray production in these
sources, advection-dominated disks are luminous radio and $\gamma$-ray
emitters, and they frequently display powerful bipolar jets of matter
escaping from the central mass, presumably containing high-energy
particles (e.g., Sambruna et al. 2004; Di Matteo et al. 2000; Allen, Di
Matteo \& Fabian 2000; Urry \& Padovani 1995; Owen, Eilek, \& Kassim
2000).

Although the effect of a standing shock in {\it heating} the gas in the
post-shock region has been examined by a number of previous authors for
both viscid (Chakrabarti \& Das 2004; Lu, Gu, \& Yuan 1999; Chakrabarti
1990) and inviscid (e.g., Lu \& Yuan 1997, 1998; Yang \& Kafatos 1995;
Abramowicz \& Chakrabarti 1990) disks, the implications of the shock for
the acceleration of {\it nonthermal} particles in the disk have not been
considered in detail before. However, a great deal of attention has been
focused on particle acceleration in the vicinity of supernova-driven
shock waves as a possible explanation for the observed cosmic-ray energy
spectrum (Blandford \& Ostriker 1978; Jones \& Ellison 1991). In the
present paper we consider the analogous process occurring in hot,
advection-dominated accretion flows (ADAFs) around black holes. These
disks are ideal sites for first-order Fermi acceleration at shocks
because the plasma is collisionless, and therefore a small fraction of
the particles can gain a great deal of energy by repeatedly crossing the
shock. Shock acceleration in the disk therefore provides an intriguing
possible explanation for the powerful outflows of relativistic particles
observed in many radio-loud systems (Le \& Becker 2004).

The idea of shock acceleration in the environment of AGNs was first
suggested by Blandford \& Ostriker (1978). Subsequently, Protheroe \&
Kazanas (1983) and Kazanas \& Ellison (1986) investigated shocks in
spherically-symmetric accretion flows as a possible explanation for the
energetic radio and $\gamma$-radiation emitted by many AGNs. However, in
these papers the acceleration of the particles was studied without the
benefit of a detailed transport equation, and the assumption of
spherical symmetry precluded the treatment of acceleration in disks. The
state of the theory was advanced significantly by Webb \& Bogdan (1987)
and Spruit (1987), who employed a transport equation to solve for the
distribution of energetic particles in a spherical accretion flow
characterized by a self-similar velocity profile terminating at a
standing shock. While more quantitative in nature than the earlier
models, these solutions are not directly applicable to disks since the
geometry is spherical and the velocity distribution is inappropriate.
Hence none of these previous models can be used to develop a single,
global, self-consistent picture for the acceleration of relativistic
particles in an accretion disk containing a shock.

Accretion disks around black holes are not spherical, except perhaps in
the innermost region, and therefore a transport equation written in
cylindrical geometry is required in order to describe the acceleration
of energetic particles in the accreting plasma. In this paper we develop
a self-consistent cylindrically-symmetric model describing the
acceleration of protons and/or electrons via first-order Fermi
acceleration processes operating in a hot, advection-dominated accretion
disk containing an isothermal shock. This is the third and final paper
in a series studying the production of relativistic particles in ADAF
disks with shocks. In Le \& Becker (2004; hereafter Paper~1), we
successfully applied our inflow/outflow model to explain the observed
kinetic power of the outflows in M87 and \SgrA, and in Le \& Becker
(2005; hereafter Paper~2), we discussed the detailed dynamics of ADAF
disks containing isothermal shocks. Based on this dynamical model we can
self-consistently compute the structure of the shocked disk and also the
rate of escape of particles and energy from the disk at the shock
location. In this paper we provide further details regarding the
distribution in space and energy of the accelerated particles in the
disk, and we also describe the energy distribution of the relativistic
particles escaping from the disk at the shock location. Our focus here
is on the computation of the Green's function for monoenergectic
particle injection, which describes in detail how the particles are
advected, diffused, and accelerated within the disk, resulting in a
nonthermal particle distribution with a high-energy power-law tail. We
also discuss the spectrum of the observable secondary radiation produced
by the escaping particles, including radio (synchrotron) as well as
inverse-Compton X-ray and $\gamma$-ray emission, which provides a
quantitative basis for developing observational tests of the model.

The first-order Fermi acceleration scenario that we focus on here (and
in Papers 1 and 2) parallels the early studies of cosmic-ray
acceleration in supernova shock waves. In analogy with the supernova
case, we expect that shock acceleration in the disk will naturally
produce a power-law energy distribution for the accelerated particles.
Our earlier work in Paper~2 established that the first-order Fermi
mechanism provides a very efficient means for accelerating the jet
particles. In this paper, the solution for the Green's function,
$\green(E,r)$, describing the evolution of monoenergetic seed particles
injected into the disk, is obtained by solving the transport equation
using the method of separation of variables. The eigenvalues and the
corresponding spatial eigenfunctions are determined using a numerical
approach because the differential equation obtained for the spatial
function cannot be solved in closed form. The resulting solution for the
Green's function provides useful physical insight into how the
relativistic particles are distributed in space and energy. We can also
use the solution for $\green(E,r)$ to check the self-consistency of the
theory by comparing the numerical values obtained for the number and
energy densities by integrating the Green's function with those obtained
in Paper~2 by directly solving the associated differential equations for
these quantities.

The remainder of the paper is organized as follows. In \S~2 we briefly
review the inviscid dynamical model used to describe the structure of
the ADAF disk/shock system. In \S~3 we develop and solve the associated
steady-state particle transport equation to obtain the solution for the
Green's function, $\green(E,r)$, describing the evolution of
monoenergetic seed particles injected from a source located at the shock
radius, and in \S~4 we evaluate the Green's function using dynamical
parameters appropriate for M87 and \SgrA. In \S~5 we discuss the
observational implications of our results for the production of
secondary radiation by the outflowing relativistic particles, and in
\S~6 we review our main conclusions.

\section{ACCRETION DYNAMICS}

It has been known for some time that inviscid accretion disks can
display both shocked and shock-free (i.e., smooth) solutions depending
on the values of the energy and angular momentum per unit mass in the
gas supplied at a large radius (e.g., Chakrabarti 1989; Chakrabarti \&
Molteni 1993; Kafatos \& Yang 1994; Lu \& Yuan 1997; Das, Chattopadhyay,
\& Chakrabarti 2001). Shocks can also exist in viscous disks if the
viscosity is relatively low (Chakrabarti 1996; Lu, Gu, \& Yuan 1999),
although smooth solutions are always possible for the same set of
upstream parameters (Narayan, Kato, \& Honma 1997; Chen, Abramowicz, \&
Lasota 1997). Hawley, Smarr, \& Wilson (1984a, 1984b) have shown through
general relativistic simulations that if the gas is falling with some
rotation, then the centrifugal force can act as a ``wall,'' triggering
the formation of a shock. Furthermore, the possibility that shock
instabilities may generate the quasi-periodic oscillations (QPOs)
observed in some sources containing black holes has been pointed out by
Chakrabarti, Acharyya, \& Molteni (2004), Lanzafame, Molteni, \&
Chakrabarti (1998), Molteni, Sponholz, \& Chakrabarti (1996), and
Chakrabarti \& Molteni (1995). Nevertheless, shocks are ``optional''
even when they are allowed, and one is always free to construct models
that avoid them. However, in general the shock solution possesses a
higher entropy content than the shock-free solution, and therefore we
argue based on the second law of thermodynamics that when possible, the
shocked solution represents the preferred mode of accretion (Becker \&
Kazanas 2001; Chakrabarti \& Molteni 1993).

The possible presence of shocks in hot ADAF disks with low viscosity
motivated us to explore in Paper~2 the relationship between the dynamical
structure of the disk/shock system and the possible acceleration of
relativistic particles that can escape from the disk to power the
outflows commonly observed from radio-loud accretion flows around both
stellar-mass and supermassive black holes. The disk/shock/outflow model
is depicted schematically in Figure~\ref{geom_fig}. In this scenario,
the gas is accelerated gravitationally toward the central mass, and
experiences a shock transition due to an obstruction near the event
horizon. The obstruction is provided by the ``centrifugal barrier,''
which is located between the inner and outer sonic points. Particles
from the high-energy tail of the background Maxwellian distribution are
accelerated at the shock discontinuity via the first-order Fermi
mechanism, resulting in the formation of a nonthermal, relativistic
particle distribution in the disk. The spatial transport of the
energetic particles within the disk is a stochastic process based on a
three-dimensional random walk through the accreting background gas.
Consequently, some of the accelerated particles diffuse to the disk
surface and become unbound, escaping through the upper and lower edges
of the cylindrical shock to form the outflow, while others diffuse
outward radially through the disk or advect across the event horizon
into the black hole.

\begfig[t]
\centering
\includegraphics[width=4.0in]{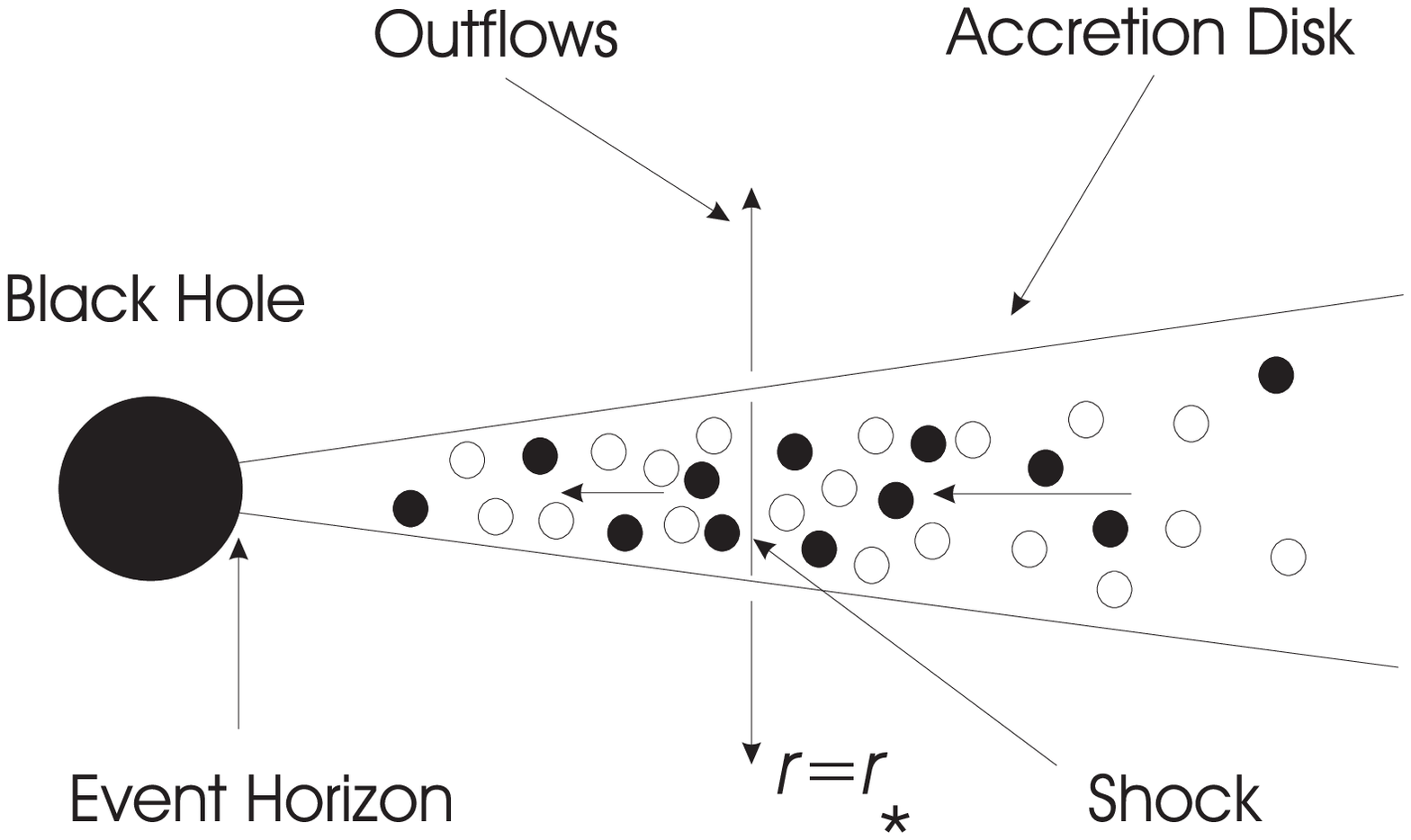}
\vskip-.1cm
\caption{Schematic representation of the disk/shock/outflow model. The
filled circles in the disk represent the accelerated particles, and the
open circles represent the MHD scattering centers which advect with the
background flow velocity. The seed particles are injected at the shock
location.}
\label{geom_fig}
\finfig

\subsection{Transport Rates}

Becker \& Le (2003) and Becker \& Subramanian (2005) demonstrated that
three integrals of the flow are conserved in viscous ADAF disks when
outflows are not occurring. The three integrals correspond to the mass
transport rate
\begeq
\dot M = 4 \pi r H \rho \, \speed \ ,
\label{eq2.1}
\fineq
the angular momentum transport rate
\begeq
\dot J = \dot M r^2 \, \Omega - \mathcal{G} \ ,
\label{eq2.2}
\fineq
and the energy transport rate
\begeq
\dot E = - \mathcal{G} \, \Omega +
\dot M\left({1 \over 2} \, v_{\phi}^2 + {1 \over 2} \, \speed^2
          + {P+U \over \rho} + \pseudophi \right) \ ,
\label{eq2.3}
\fineq
where $\rho$ is the mass density, $\Omega$ is the angular velocity,
${\cal G}$ is the torque, $H$ is the disk half-thickness, $\speed$ is
the positive radial inflow speed, $v_{\phi}=r \, \Omega$ is the
azimuthal velocity, $U$ is the internal energy density, $P =
(\gamma-1)\,U$ is the pressure, and
\begeq
\pseudophi(r) \equiv {-GM \over r-\rs}
\label{eq2.4}
\fineq
denotes the pseudo-Newtonian gravitational potential for a black hole
with mass $M$ and Schwarzschild radius $\rs=2GM/c^2$ (Paczy\'nski \&
Wiita 1980). Each of the various densities and velocities represents a
vertical average over the disk structure. We also assume that the ratio
of specific heats, $\gamma$, maintains a constant value throughout the
flow. Note that all of the transport rates $\dot M$, $\dot J$, and $\dot
E$ are defined to be positive for inflow.

The quantities $\dot M$ and $\dot J$ remain constant throughout the
entire flow, and therefore they represent the rates at which mass and
angular momentum, respectively, enter the black hole. The energy
transport rate $\dot E$ also remains constant, except at the location of
an isothermal shock if one is present in the disk. The torque ${\cal G}$
is related to the gradient of the angular velocity $\Omega$ via the
usual expression (e.g., Frank et al. 2002)
\begeq
\mathcal{G} =
- 4 \pi r^3 H \rho \, \nu {d\Omega \over dr} \ ,
\label{eq2.5}
\fineq
where $\nu$ is the kinematic viscosity. The disk half-thickness $H$ is
given by the standard hydrostatic prescription
\begeq
H(r) = {c_s \over \OmegaK} \ ,
\label{eq2.6}
\fineq
where $c_s$ represents the adiabatic sound speed, defined by
\begeq
c_s(r) \equiv \left({\gamma P \over \rho}\right)^{1/2}  \ ,
\label{eq2.7}
\fineq
and $\OmegaK$ denotes the Keplerian angular velocity of matter
in a circular orbit at radius $r$ in the pseudo-Newtonian potential
(eq.~[\ref{eq2.4}]), defined by
\begeq
\OmegaK^2(r) \equiv {GM \over r(r-\rs)^2}
= {1 \over r} {d\pseudophi \over dr} \ .
\label{eq2.8}
\fineq

\subsection{Inviscid Flow Equations}

Following the approach taken in Paper~2, we focus here on the inviscid
case since that allows the disk structure to be computed analytically,
while retaining reasonable agreement with the dynamics of actual disks
with low viscosities. We utilize the set of physical conservation
equations employed by Chakrabarti (1989) and Abramowicz \& Chakrabarti
(1990), who investigated the structure of one-dimensional, inviscid,
steady-state, axisymmetric accretion flows. In the inviscid limit, the
torque ${\cal G}$ vanishes, and the angular momentum per unit mass
transported through the disk is given by
\begeq
\ell(r) \equiv {\dot J \over \dot M} = r^2 \, \Omega(r) = {\rm constant} \ .
\label{eq2.9}
\fineq
We can combine equations~(\ref{eq2.3}) and (\ref{eq2.7}) to show that
the energy transported per unit mass in the inviscid case can be written
as
\begeq
\epsilon \equiv {\dot E \over \dot M} =
{1 \over 2} \, \speed^2 + {1 \over 2} \, {\ell^2
\over r^2} + {c_s^2 \over \gamma-1} + \pseudophi \ ,
\label{eq2.10}
\fineq
where we have also utilized the relation $\ell = r v_\phi$. The
resulting disk/shock model depends on three free parameters, namely the
energy transport per unit mass $\epsilon$, the specific heats ratio
$\gamma$, and the specific angular momentum $\ell$. The value of
$\epsilon$ will jump at the location of an isothermal shock if one
exists in the disk, but the value of $\ell$ remains constant throughout
the flow. This implies that the specific angular momentum of the
particles escaping through the upper and lower surfaces of the
cylindrical shock must be equal to the average value of the specific
angular momentum for the particles remaining in the disk, and therefore
the outflow exerts no torque on the disk (Becker, Subramanian, \&
Kazanas 2001).

The values of the energy transport parameter $\epsilon$ on the upstream
and downstream sides of the isothermal shock at $r=r_*$ are denoted by
$\epsilon_-$ and $\epsilon_+$, respectively. The condition $\epsilon_- >
\epsilon_+$ must be satisfied as a consequence of the loss of energy
through the upper and lower surfaces of the disk at the shock location.
In order to satisfy global energy conservation, the rate at which energy
escapes from the gas as it crosses the shock must be equal to the jet
luminosity, $\Ljet$, and therefore we have
\begeq
\Ljet \equiv -\Delta \dot E = -\dot M \, \Delta \epsilon
\ \ \propto \ {\rm ergs \ s}^{-1} \ ,
\label{eq2.11}
\fineq
where $\Delta$ denotes the difference between quantities on the
downstream and upstream sides of the shock, so that $\Delta\epsilon =
\epsilon_+ -\epsilon_- < 0$. The drop in $\epsilon$ at the shock has
the effect of altering the transonic structure of the flow in the
post-shock region, compared with the shock-free configuration (see
Paper~2). In the inviscid case, the flow is purely adiabatic except at
the location of an isothermal shock (if one is present), and therefore
the pressure is related to the density profile via
\begeq
P = D_0 \, \rho^\gamma \ ,
\label{eq2.12}
\fineq
where $D_0$ is a parameter related to the specific entropy that remains
constant except at the location of a shock. The process of constructing
a dynamical model for a specific disk/shock system begins with the
selection of values for the fundamental parameters $\epsilon_-$, $\ell$,
and $\gamma$. For given values of these three quantities, a unique flow
structure can be obtained using the procedure discussed in \S\S~3 and 4
of Paper~2, which yields a numerical solution for the inflow speed
$\speed(r)$. Following Narayan, Kato, \& Honma (1997), we shall assume
an approximate equipartition between the gas and magnetic pressures, and
therefore we set $\gamma=1.5$. The range of acceptable values for
$\epsilon_-$ and $\ell$ is constrained by the observations of a specific
object, as discussed in \S~7 of Paper~2.

\section{STEADY-STATE PARTICLE TRANSPORT}

The gas in ADAF disks is hot and tenuous, and therefore essentially
collisionless. In this situation, the relativistic particles diffuse
mainly via collisions with magnetohydrodynamical (MHD) scattering
centers (e.g., Alfv\'en waves), which control both the spatial transport
and the acceleration of the particles. The magnetic scattering centers
are advected along with the background flow, and therefore the shock
width is expected to be comparable to the magnetic coherence length,
$\lambda_{\rm mag}$. Particles from the high-energy tail of the
background Maxwellian distribution are accelerated at the shock
discontinuity via the first-order Fermi mechanism, resulting in the
formation of a nonthermal, relativistic particle distribution in the
disk. The probability of multiple shock crossings decreases
exponentially with the number of crossings, and the mean energy of the
particles increases exponentially with the number of crossings. This
combination of factors naturally gives rise to a power-law energy
distribution, which is a general characteristic of Fermi processes
(Fermi 1954).

Two effects limit the maximum energy that can be achieved by the
particles. First, at very high energies the particles will tend to lose
energy to the waves due to recoil. Second, the mean free path
$\lambda_{\rm mag}$ will eventually exceed the thickness of the disk as
the particle energy is increased, resulting in escape from the disk
without further acceleration. As a consequence of the random walk, some
of the accelerated particles diffuse to the disk surface and become
unbound, escaping through the upper and lower edges of the cylindrical
shock to form the outflow, while others diffuse outward radially through
the disk or advect across the event horizon into the black hole (see
Fig.~\ref{geom_fig}). Our goal in this paper is to analyze the Green's
function $\green(E,r)$ in the disk based on the transport equation and
the dynamical results discussed in Paper~2. We focus specifically on
models~2 and 5 from Paper~2, which describe the disk structures for M87
and \SgrA, respectively.

\subsection{Transport Equation}

The particle transport formalism we utilize includes advection, spatial
diffusion, first-order Fermi acceleration, and particle escape. Our
treatment of the Fermi process includes both the general compression
related to the overall convergence of the accretion flow, as well as the
enhanced compression that occurs at the shock. To avoid unnecessary
mathematical complexity, we utilize a simplified model for the spatial
transport in which only the radial ($r$) component is treated in detail.
In this approach, the diffusion and escape of the particles in the
vertical ($z$) direction is modeled using an escape-probability
formalism. We will adopt the test-particle approximation utilized in
Paper~2, meaning that the dynamical effect of the relativistic particle
pressure on the flow structure is assumed to be negligible. This
assumption is valid provided the pressure of the relativistic particles
is much smaller than the background (thermal) pressure, which is
verified for models~2 and 5 in \S~8 of Paper~2. We also assume that the
injection of the seed particles and the escape of the accelerated
particles from the disk are both localized at the isothermal shock
radius, which is consistent with the strong concentration of first-order
Fermi acceleration in the vicinity of the shock.

The localization of the particle injection and escape at the shock
radius allows us to make a direct connection between the jump in the
energy flux in the disk at the shock location and the energy carried
away by the escaping particles. This connection is essential in order to
maintain self-consistency between the dynamical and transport
calculations, as discussed in Paper~2. In a steady-state situation, the
Green's function, $\green(E,r)$, representing the particle distribution
resulting from the continual injection of monoenergetic seed particles
from a source located at the shock radius ($r=r_*$) satisfies the
transport equation (Becker 1992)
\begeq
{\partial \green \over \partial t} = 0 = -\vec\nabla \cdot
\vec F - {1 \over 3 E^2} {\partial \over \partial E} \left(E^3 \,
\vec v \cdot \vec\nabla \green \right) + \dot f_{\rm source} - \dot
f_{\rm esc} \ ,
\label{eq3.1}
\fineq
where the specific flux $\vec F$ is evaluated using
\begeq
\vec F = -\kappa \vec\nabla\green - \vec v \, {E \over 3}
\, {\partial \green \over \partial E} \ ,
\label{eq3.2}
\fineq
and the quantities $E$, $\kappa$, and $\vec v$ represent the particle
energy, the spatial diffusion coefficient, and the vector velocity,
respectively. The velocity has components given by $\vec v = v_r \hat r
+ v_z \hat z + v_\phi \, \hat\phi$, with $v_r = - \speed < 0$ since we
are dealing with inflow.

The injection of the monoenergetic seed particles is described by the
function
\begeq
\dot f_{\rm source} = {\N0 \, \delta(E-E_0) \, \delta(r-r_*)
\over (4 \pi E_0)^2 \, r_* \, H_*} \ ,
\label{eq3.3}
\fineq
where $E_0$ is the energy of the injected particles, $\dot N_0$ is the
particle injection rate, and $H_* \equiv H(r_*)$ represents the
half-thickness of the disk at the shock location. The escape of the
particles through the upper and lower surfaces of the disk is
represented by the term
\begeq
\dot f_{\rm esc} = A_0 \, c \, \delta(r-r_*) \, \green \ ,
\label{eq3.4}
\fineq
where the dimensionless parameter $A_0$ is computed using
(see Appendix~B of Paper~2)
\begeq
A_0 = \left(3 \kappa_* \over c H_*\right)^2 < 1 \ ,
\label{eq3.5}
\fineq
and $\kappa_* \equiv (\kappa_-+\kappa_+)/2$ denotes the mean value of
the spatial diffusion coefficient at the shock radius. The subscripts
``-'' and ``+'' will be used to denote quantities measured just upstream
and just downstream from the shock, respectively. The condition $A_0 <
1$ is required for the validity of the diffusive treatment of the
vertical escape employed in our approach.

In order to compute the Green's function $\green(E,r)$, we must specify
the injection energy of the seed particles $E_0$ as well as their
injection rate $\dot N_0$. Following the approach taken in Paper~2, we set
the injection energy using $E_0 = 0.002\,$ergs, which corresponds to an
injected Lorentz factor $\Gamma_0 \equiv E_0 / (m_p \, c^2) \sim 1.3$,
where $m_p$ is the proton mass. Particles injected with energy $E_0$ are
subsequently accelerated to much higher energies due to repeated shock
crossings. We find that the speed of the injected particles, $v_0 = c \,
(1-\Gamma_0^{-2})^{1/2}$, is several times larger than the mean ion
thermal velocity at the shock location, $v_{\rm rms} = (3 k
T_*/m_p)^{1/2}$, where $T_*$ is the ion temperature at the shock. The
seed particles are therefore picked up from the high-energy tail of the
local Maxwellian distribution. With $E_0$ specified, we can compute the
particle injection rate $\N0$ using the energy conservation condition
\begeq
\dot N_0 \, E_0 = - \dot M \, \Delta \epsilon
\ \ \propto \ {\rm ergs \ s}^{-1} \ ,
\label{eq3.6}
\fineq
where $\Delta \epsilon < 0$ represents the jump in the energy transport
per unit mass as the plasma crosses the shock (see eqs.~[\ref{eq2.10}]
and [\ref{eq2.11}]). This relation ensures that the energy injection
rate for the seed particles is equal to the energy loss rate for the
background gas at the isothermal shock location.

The total number and energy densities of the relativistic particles,
denoted by $n_r$ and $U_r$, respectively, are related to the Green's
function via
\begeq
n_r(r) = \int_0^\infty 4 \pi E^2 \, \green(E,r) \, dE \ , \ \ \ \
\ U_r(r) = \int_0^\infty 4 \pi E^3 \, \green(E,r) \, dE \ ,
\label{eq3.7}
\fineq
which determine the normalization of $\green$. Equations~(\ref{eq3.1})
and (\ref{eq3.2}) can be combined to obtain the alternative form
\begeq
\vec v \cdot \vec\nabla \green = {E \over 3} \, {\partial \green
\over \partial E} \, \vec\nabla \cdot \vec v
+ \vec\nabla
\cdot \left(\kappa \, \vec\nabla \green \right) + \dot f_{\rm
source} - \dot f_{\rm esc} \ ,
\label{eq3.8}
\fineq
where the left-hand side represents the co-moving (advective) time
derivative and the terms on the right-hand side describe first-order
Fermi acceleration, spatial diffusion, the particle source, and the
escape of particles from the disk at the shock location, respectively.
Note that escape and particle injection are localized to the shock
radius due to the presence of the $\delta$-functions in
equations~(\ref{eq3.3}) and (\ref{eq3.4}). Our focus here is on the
first-order Fermi acceleration of relativistic particles at a standing
shock in an accretion disk, and therefore equation~(\ref{eq3.8}) does
not include second-order Fermi processes that may also occur in the flow
due to MHD turbulence (e.g., Schlickeiser 1989a,b; Subramanian, Becker,
\& Kazanas 1999; Becker et al. 2006). Since no energy loss mechanisms
are included in the model, all of the particles injected with energy
$E_0$ will experience an increase in energy.

Under the assumption of cylindrical symmetry, equations~(\ref{eq3.3}),
(\ref{eq3.4}), and (\ref{eq3.8}) can be combined to obtain
\begeqarray
v_r {\partial \green \over \partial r} + v_z {\partial
\green \over \partial z} &-&
{E \over 3} \, {\partial \green \over \partial E}
\left[{1 \over r} {d \over d r} \left(r v_r\right)
+ {d v_z \over dz}\right]
- {1 \over r} {\partial \over
\partial r} \left(r \kappa {\partial \green \over \partial r}\right)
\nonumber \\
&=& {\N0 \, \delta(E-E_0) \, \delta(r-r_*) \over (4 \pi E_0)^2 \,
r_* H_*} - A_0 \, c \, \delta(r-r_*) \, \green \ ,
\label{eq3.9}
\fineqarray
where the escape of particles from the disk is described by the final
term on the right-hand side. We focus here on the vertically-integrated
form of the transport equation, which can be written as (see Paper~2,
Appendix~A)
\begeqarray
- H \speed {\partial \green \over \partial r} &=& - {1 \over r}
\, {d \over dr} (r H \speed) \, {E \over 3} \, {\partial \green
\over \partial E} + {1 \over r} {\partial \over \partial r}
\left(r H \kappa \, {\partial \green \over \partial r}\right)
\nonumber \\
&+& {\dot N_0 \, \delta(E-E_0) \, \delta(r-r_*) \over (4 \pi E_0)^2 r_*}
- A_0 \, c \, H_* \, \delta(r-r_*) \, \green \ ,
\label{eq3.10}
\fineqarray
where $\speed = -v_r$ and $\green$ and $\kappa$ are vertically averaged
quantities. Due to the presence of the velocity derivative on the
right-hand side of equation~(\ref{eq3.10}), the first-order Fermi
acceleration of the particles is most pronounced near the shock, where
$u$ is discontinuous. In the vicinity of the shock, we find that
\begeq
{d\speed \over dr} \to (\speed_- - \speed_+) \, \delta(r-r_*) \ ,
\ \ \ \ \ r \to r_* \ ,
\label{eq3.11}
\fineq
where $\speed_-$ and $\speed_+$ denote the positive inflow speeds just
upstream and downstream from the shock, respectively. The velocity jump
condition appropriate for an isothermal shock is (see Paper~2)
\begeq
{\speed_+ \over \speed_-}
= {1 \over \gamma \, {\cal M}_-^2} \ ,
\label{eq3.12}
\fineq
where ${\cal M}_-$ is the incident Mach number on the upstream side of
the shock. Although the Fermi acceleration of the particles is
concentrated at the shock, the rest of the disk also contributes to the
particle acceleration because of the general convergence of the MHD
waves in the accretion flow.

In order to close the system of equations and solve for the
Green's function using equation~(\ref{eq3.10}), we must also specify
the radial variation of the diffusion coefficient $\kappa$. The
behavior of $\kappa$ can be constrained by considering the
fundamental physical principles governing accretion onto a black
hole, which we have demonstrated in Paper~2. The precise
functional form for the spatial variation of $\kappa$ is not
completely understood in the accretion disk environment. In order
to obtain a mathematically tractable set of equations with a
reasonable physical behavior, we utilize the general form
(see Paper~2, \S~5)
\begeq
\kappa(r) = \kappa_0 \, \speed(r) \, \rs \left({r \over \rs} -1
\right)^2 \ ,
\label{eq3.13}
\fineq
where $\kappa_0 > 0$ is a dimensionless constant that can be computed
for a given source based on energy considerations (see \S~4). Due to the
appearance of the radial speed $\speed$ in equation~(\ref{eq3.13}), we
note that $\kappa$ exhibits a jump at the shock. This is expected on
physical grounds since the MHD waves that scatter the ions are swept
along with the thermal background flow, and therefore they should also
experience a density compression at the shock. We also point out that
the dimensionless escape parameter $A_0$ can be evaluated by combining
equations~(\ref{eq3.5}) and (\ref{eq3.13}), which yields
\begeq
A_0 = \left({3 \kappa_0 \speed_* \rs \over c H_*}\right)^2
\left({r_* \over \rs} - 1\right)^4 < \, 1 \ ,
\label{eq3.14}
\fineq
where $\speed_* \equiv (\speed_- + \speed_+)/2$ denotes the mean value
of the inflow speed at the shock location. Note that the value of the
diffusion parameter $\kappa_0$ is constrained by the inequality in
equation~(\ref{eq3.14}).

\subsection{Separation of Variables}

For values of the particle energy $E > E_0$, the source term in
equation~(\ref{eq3.10}) vanishes and the remaining equation is separable
in energy and space using the functions
\begeq
f_{_n}(E,r) = \left(E \over E_0\right)^{-\lambda_n} \, Y_n(r) \ ,
\label{eq3.15}
\fineq
where $\lambda_n$ are the eigenvalues, and the spatial eigenfunctions
$Y_n(r)$ satisfy the second-order ordinary differential equation
\begeq
- H \speed {d Y_n \over d r} = {\lambda_n \over 3 r}
{d \over dr} (r H \speed) \, Y_n + {1 \over r} {d \over dr}
\left(r H \kappa \, {d Y_n \over d r}\right)
- A_0 \, c \, H_* \, \delta(r-r_*) \, Y_n \ .
\label{eq3.16}
\fineq
The eigenvalues $\lambda_n$ are determined by applying suitable boundary
conditions to the spatial eigenfunctions, as discussed below. Once a
numerical solution for the inflow speed $\speed(r)$ has been obtained
following the procedures described in Paper~2, we can compute $H(r)$ and
$\kappa(r)$ using equations~(\ref{eq2.6}) and (\ref{eq3.13}),
respectively. Since the coefficients in equation~(\ref{eq3.16}) cannot
be represented in closed form, analytical solutions cannot be obtained
and therefore the eigenfunctions $Y_n(r)$ must be determined
numerically. Away from the shock ($r \ne r_*$), equation~(\ref{eq3.16})
reduces to
\begeq
{d^2 Y_n \over dr^2} + \left[{d\ln(r H \kappa) \over dr}
+ {\speed \over \kappa}\right] {d Y_n \over dr} +
{\lambda_n \speed Y_n \over 3 \, \kappa} \, {d \ln(r H \speed)\over dr}
= 0 \ ,
\label{eq3.17}
\fineq
which can be rewritten as
\begeq
{d^2 Y_n \over dr^2} + \left[{d\ln(r H \speed) \over dr}
+ {2 \over r-\rs} + {\rs \over \kappa_0 (r-\rs)^2}\right]
{d Y_n \over dr} + {\lambda_n \rs Y_n \over 3 \kappa_0
(r-\rs)^2} \, {d \ln(r H \speed)\over dr} = 0 \ ,
\label{eq3.18}
\fineq
where we have used equation~(\ref{eq3.13}) to substitute for $\kappa$.
Once a solution for the inflow speed $\speed(r)$ has been determined
using the procedure described in \S~2, the variation of the disk
half-thickness $H(r)$ can be computed by combining
equations~(\ref{eq2.1}), (\ref{eq2.6}), (\ref{eq2.7}), and
(\ref{eq2.12}). The result obtained is
\begeq
H(r) = {1 \over \OmegaK(r)}
\left[r^{3/2} (r-\rs) \, \speed(r) \over K\right]^{(1-\gamma)/(1+\gamma)}
\ ,
\label{eq3.19}
\fineq
where $\OmegaK$ is given by equation~(\ref{eq2.8}) and $K$ is the
``entropy parameter'', which remains constant except at the location of
an isothermal shock if one is present in the flow (see Paper~2 and
Becker \& Le 2003). By utilizing the solutions obtained for $\speed(r)$
and $H(r)$, we are able to compute all of the coefficients in
equation~(\ref{eq3.18}), which allows us to solve numerically for the
spatial eigenfunctions $Y_n(r)$.

\subsection{Eigenvalues and Eigenfunctions}

The global solution for the eigenfunction $Y_n(r)$ must satisfy
continuity and derivative jump conditions associated with the presence
of the shock/source at radius $r=r_*$. By integrating
equation~(\ref{eq3.16}) with respect to radius in the vicinity of the
shock, we find that
\begeqarray \Delta [Y_n] & = & 0 \ ,
\label{eq3.20} \\
\Delta \left[\left({\lambda_n \over 3} \, \speed \, Y_n
+ \kappa \, {d Y_n \over dr} \right) \right] & = & - c A_0
Y_n(r_*)
\label{eq3.21} \ ,
\fineqarray
where the symbol $\Delta$ denotes the difference between postshock and
preshock quantities, and we have made use of the fact that $H$ is
continuous across the isothermal shock. Equations~(\ref{eq3.20}) and
(\ref{eq3.21}) establish that $Y_n(r)$ must be continuous at the shock
location, and its derivative must display a jump there.

Since the spatial eigenfunctions $Y_n(r)$ satisfy the second-order
linear differential equation~(\ref{eq3.18}), we must also impose two
boundary conditions in order to determine the global solutions and the
associated eigenvalues $\lambda_n$. The required conditions can be
developed based on our knowledge of the dynamical structure of the disk
near the event horizon ($r = \rs$) and at large radii (see Becker \& Le
2003 and \S~4 of Paper~2). By utilizing the Frobenius expansion method
to identify the dominant terms in equation~(\ref{eq3.18}), we find that
the global solutions for the spatial eigenfunctions can be written as
\begeq Y_n(r) = \left\{\begin{array}{ll}
      G^{\rm in}_n(r)  & , \ \  r \le r_* \ , \nonumber \\
      \nonumber \\
      a_n G^{\rm out}_n(r) & , \ \  r \ge r_* \ ,
                   \end{array} \right.
\label{eq3.22}
\fineq
where $a_n$ is a matching constant and $G^{\rm in}_n(r)$ and $G^{\rm
out}_n(r)$ denote the fundamental solutions to equation~(\ref{eq3.18})
possessing the asymptotic behaviors
\begeq
\begin{array}{ll}
G^{\rm in}_n(r) \to \ g^{\rm in}_n(r) \equiv \left({r \over \rs}-1
\right)^{-\lambda_n/(3\gamma+3)}
& , \ \ r \to \rs \ , \ \ \nonumber \\
\nonumber \\
G^{\rm out}_n(r) \to \ g^{\rm out}_n(r) \equiv \left({r \over \rs}
\right)^{-1} & , \ \ r \to \infty \ ,
\end{array}
\label{eq3.23}
\fineq
at small and large radii, respectively. The value of $a_n$ is determined
using the continuity condition (see eq.~[\ref{eq3.20}]), which yields
\begeq
a_n = {G^{\rm in}_n(r_*) \over G^{\rm out}_n(r_*)} \ .
\label{eq3.24}
\fineq
The validity of the asymptotic forms in equations~(\ref{eq3.23}) is
verified in \S~4.1 via comparison with the numerical solutions obtained
for the spatial eigenfunctions.

We use a bi-directional integration scheme to solve for the eigenvalues
$\lambda_n$ based on the boundary conditions in
equations~(\ref{eq3.23}). The value of $\lambda_n$ is iterated using a
bisection method until the Wronskian of the inner and outer solutions
vanishes at a matching radius located in the postshock region. Once a
particular eigenvalue has been determined to the required accuracy, the
matching constant $a_n$ is computed using equation~(\ref{eq3.24}). This
procedure is repeated until the desired number of eigenvalues and
eigenfunctions have been obtained. The sequences of eigenvalues
associated with the model~2 (M87) and model~5 (\SgrA) parameters are
plotted in Figure~\ref{eigenfig}. Note that $\lambda_1 \sim 4$ in all
cases, indicating that the acceleration is quite efficient, in analogy
with the case of cosmic-ray acceleration (e.g., Blandford \& Ostriker
1978). The first two eigenvalues in models~2 and 5 are
$\lambda_1=4.165$, $\lambda_2=6.415$ and $\lambda_1=4.180$,
$\lambda_2=6.344$, respectively. It follows that at high energies the
power-law slope of the energy distribution is dominated by the first
eigenvalue $\lambda_1$, since all of the subsequent eigenvalues are much
larger (see Fig.~\ref{eigenfig}).

\begin{figure}[t]
\vskip-12.5cm
\hskip 1.0cm
\includegraphics[width=5.5in]{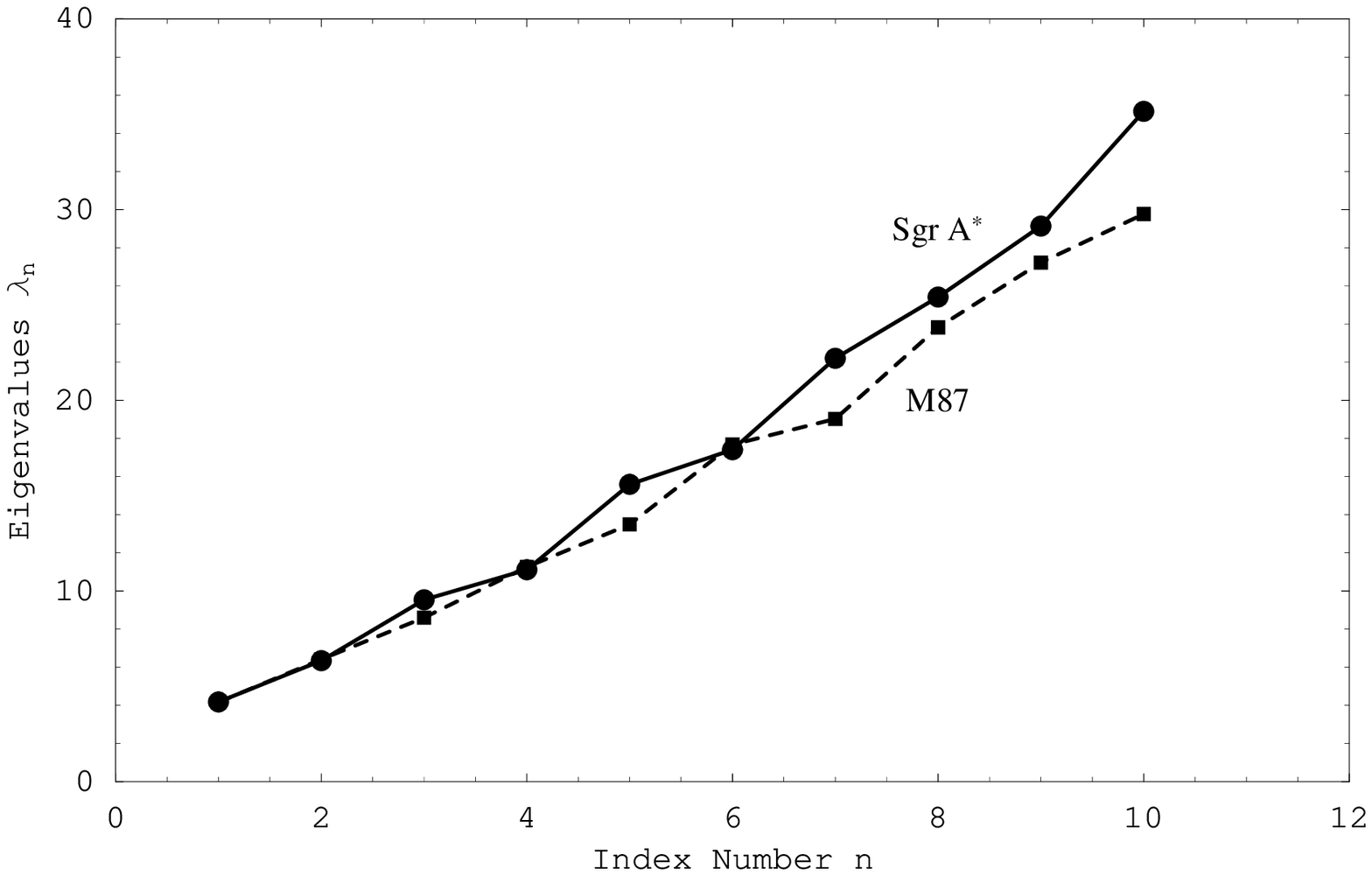}
\vskip-12.5cm
\caption{Eigenvalues for model~2 (M87; filled
squares) and model~5 (\SgrA; filled circles).}
\label{eigenfig}
\finfig

\subsection{Eigenfunction Orthogonality}

We can establish several useful general properties of the eigenfunctions
by rewriting equation~(\ref{eq3.18}) in the Sturm-Liouville form
\begeq
{d \over dr}\left[S(r) \, {dY_n \over dr}\right]
+ \lambda_n \, \omega(r) \, Y_n(r) = 0 \ ,
\label{eq3.25}
\fineq
where
\begeq
S(r) \equiv {r H \kappa \over r_* H_* \kappa_*}
\ \exp\left\{{1 \over \kappa_0}\left[\left({r_* \over \rs}-1\right)^{-1}
- \left({r \over \rs}-1\right)^{-1} \right]\right\}
\ ,
\label{eq3.26}
\fineq
and the weight function $\omega(r)$ is defined by
\begeq
\omega(r) \equiv {\speed \, S \over 3 \, \kappa} \, {d \ln(r H \speed)
\over dr}
\ .
\label{eq3.27}
\fineq
Note that $\omega(r)$ displays a $\delta$-function discontinuity at the
shock through its dependence on the derivative of $\speed(r)$. In the
vicinity of the shock, we can combine equations~(\ref{eq3.11}),
(\ref{eq3.26}), and (\ref{eq3.27}) to show that
\begeq
\omega(r) \to \left({\speed_- - \speed_+ \over 3 \kappa_*}\right)
\delta(r-r_*) \ , \ \ \ \ \ r \to r_* \ .
\label{eq3.28}
\fineq
Based on the boundary conditions satisfied by the spatial eigenfunctions
$Y_n(r)$ (see eqs.~[\ref{eq3.23}]), we demonstrate in Appendix~A that
the eigenfunctions satisfy the orthogonality relation
\begeq
\int^{\infty}_{\rs} Y_n(r) \, Y_m(r) \, \omega(r) \, dr = 0 \ ,
\ \ \ \ \ m \neq n \ .
\label{eq3.29}
\fineq

\subsection{Eigenfunction Expansion}

We have established that $Y_n(r)$ is the solution to a standard
Sturm-Liouville problem, and therefore it follows that the set of
eigenfunctions is complete. Consequently, the Green's function
$\green(E,r)$ can be represented using the eigenfunction expansion (see
eq.~[\ref{eq3.15}])
\begeq
\green(E,r) = \sum^{N_{\rm max}}_{n=1} b_n Y_n(r)
\left(E \over E_0\right)^{-\lambda_{\rm n}} \ ,
\ \ \ \ \ E \ge E_0 \ ,
\label{eq3.30}
\fineq
where $b_n$ are the expansion coefficients (with either positive
or negative signs) and the value of $N_{\rm max}$ is established by
analyzing the term-by-term convergence of the series. We utilize the
value $N_{\rm max}=10$ in our numerical examples, which generally yields
an accuracy of at least three decimal digits. Note that $\green(E,r)=0$
for all $E < E_0$ because no deceleration processes are included in the
particle transport model.

The expansion coefficients $b_1, b_2, b_3, \ldots$ can be calculated
by utilizing the orthogonality of the spatial eigenfunctions. At the
source energy, $E=E_0$, equation~(\ref{eq3.30}) reduces to
\begeq
\green(E_0,r) =  \sum^{N_{\rm max}}_{m=1}
b_m \, Y_m(r) \ .
\label{eq3.31}
\fineq
Multiplying each side of this expression by the product $Y_n(r) \,
\omega(r)$ and integrating from $r=\rs$ to $r=\infty$ yields
\begeq
\int^{\infty}_{\rs} \green(E_0,r) \, Y_n(r) \, \omega(r) \, dr
= \sum^{N_{\rm max}}_{m=1} b_m \int^{\infty}_{\rs} Y_m(r) \
Y_n(r) \, \omega(r) \, dr \ .
\label{eq3.32}
\fineq
Based on the orthogonality of the eigenfunctions, as expressed by
equation~(\ref{eq3.29}), we note that only the $m=n$ term survives on
the right-hand side of equation~(\ref{eq3.32}), and we therefore obtain
\begeq
\int^{\infty}_{\rs} \green(E_0,r) \, Y_n(r) \, \omega(r) \, dr
= b_n \int^{\infty}_{\rs} Y^2_n(r) \, \omega(r) \, dr \ .
\label{eq3.33}
\fineq
Solving this expression for the expansion coefficient $b_n$ yields
\begeq
b_n = {\int^\infty_{\rs} \green(E_0,r) \, Y_n(r) \, \omega(r)
\, dr \over {\cal I}_n} \ ,
\label{eq3.34}
\fineq
where the quadratic normalization integrals, ${\cal I}_n$, are
defined by
\begeq
{\cal I}_n \equiv \int^\infty_{\rs} Y^2_n(r) \, \omega(r) \, dr \ .
\label{eq3.35}
\fineq

To complete the calculation of the expansion coefficients, we
need to evaluate the distribution function at the source energy,
$\green(E_0,r)$. This can be accomplished by using
equation~(\ref{eq3.11}) to substitute for the velocity derivative in
equation~(\ref{eq3.10}) and then integrating with respect to $E$ in a
small range around the injection energy $E_0$, which yields
\begeq
\green(E_0,r) = \left\{
\begin{array}{ll}
\displaystyle{3 \dot N_0 \over (4 \pi)^2 E_0^3 \, r_* H_*
(\speed_ - -\speed_+)} \ , & r = r_* \ , \\[12pt]
\displaystyle{0} \ , & r \ne r_* \ .
\end{array}
\right.
\label{eq3.36}
\fineq
Although the source term in the transport equation~(\ref{eq3.9}) is a
delta function, we note that the Green's function evaluated at the
injection energy $E_0$ is {\it not} a delta function. Instead, as
indicated by equation~(\ref{eq3.36}), we find that the Green's function
at the injection energy is zero for radii away from the shock, and it is
equal to a finite value at the shock ($r=r_*$). From a physical point of
view, this reflects the fact that the particle injection occurs at the
shock location in our model. The strong acceleration experienced by the
particles at the shock eliminates the appearance of a delta-function
behavior in the resulting Green's function. It is interesting to compare
this with the behavior observed in the case of particle acceleration in
a plane-parallel supernova shock wave, which was analyzed in detail by
Blandford \& Ostriker (1978). Our results are similar to theirs, except
that the Blandford \& Ostriker Green's function evaluated at the
injection energy $E_0$ has a finite value at {\it all} locations in the
flow, because there is no convergence in the plasma surrounding the
plane-parallel supernova shock wave. However, in our model, the plasma
converges at all radii in the accretion flow (not just at the shock),
and this additional particle acceleration causes the Green's function to
vanish at the injection energy for all $r \ne r_*$.

Utilizing equation~(\ref{eq3.36}) to substitute for $\green(E_0,r)$ in
equation~(\ref{eq3.34}) and carrying out the integration, we obtain
\begeq
b_n = {\dot N_0 \, Y_n(r_*) \over (4 \pi)^2 E_0^3 \, r_*
H_* \kappa_* {\cal I}_n} \ ,
\label{eq3.37}
\fineq
where we have made use of the fact that the weight function $\omega(r)$
has a $\delta$-function behavior close to the shock (see
eq.~[\ref{eq3.28}]). The singular nature of the weight function also
needs to be considered when computing the normalization integrals ${\cal
I}_n$ defined in equation~(\ref{eq3.35}). By combining
equations~(\ref{eq3.28}) and (\ref{eq3.35}), we find that
\begeq
{\cal I}_n = \lim_{\epsilon \to 0}
\int_{\rs}^{r_{*} - \epsilon} \omega(r) \, Y^2_n(r) \, dr
+ \int_{r_{*} + \epsilon}^{\infty} \omega(r) \, Y^2_n(r) \, dr
+ \left(\speed_- - \speed_+ \over 3 \kappa_*\right)
Y_n^2(r_*) \ ,
\label{eq3.38}
\fineq
and we shall use this expression to evaluate the normalization integrals.

\section{ASTROPHYSICAL APPLICATIONS}

In Paper~2 we investigated the acceleration of particles in an inviscid
ADAF disk containing an isothermal shock. For a given source with a
known black hole mass $M$, accretion rate $\dot M$, and jet kinetic
power $\Ljet$, we found that a family of flow solutions can be obtained,
with each solution corresponding to a different value of the diffusion
parameter $\kappa_0$ (see eq.~[\ref{eq3.13}]). We concluded that the
upstream energy flux, $\epsilon_-$, is maximized for a certain value of
$\kappa_0$, and we adopted that particular value for $\kappa_0$ in our
models for M87 and \SgrA~in Paper~2. In Table~\ref{tbl-1} we list the
values of the parameters $M$, $\dot M$, $\Ljet$, $\epsilon_-$, $\ell$,
$\epsilon_+$, $r_*$, and $\speed_*$ associated with models~2 and 5,
which describe M87 and \SgrA, respectively.

\begfig[t]
\vskip-0cm \hskip-2.0cm
\includegraphics[width=7.0in]{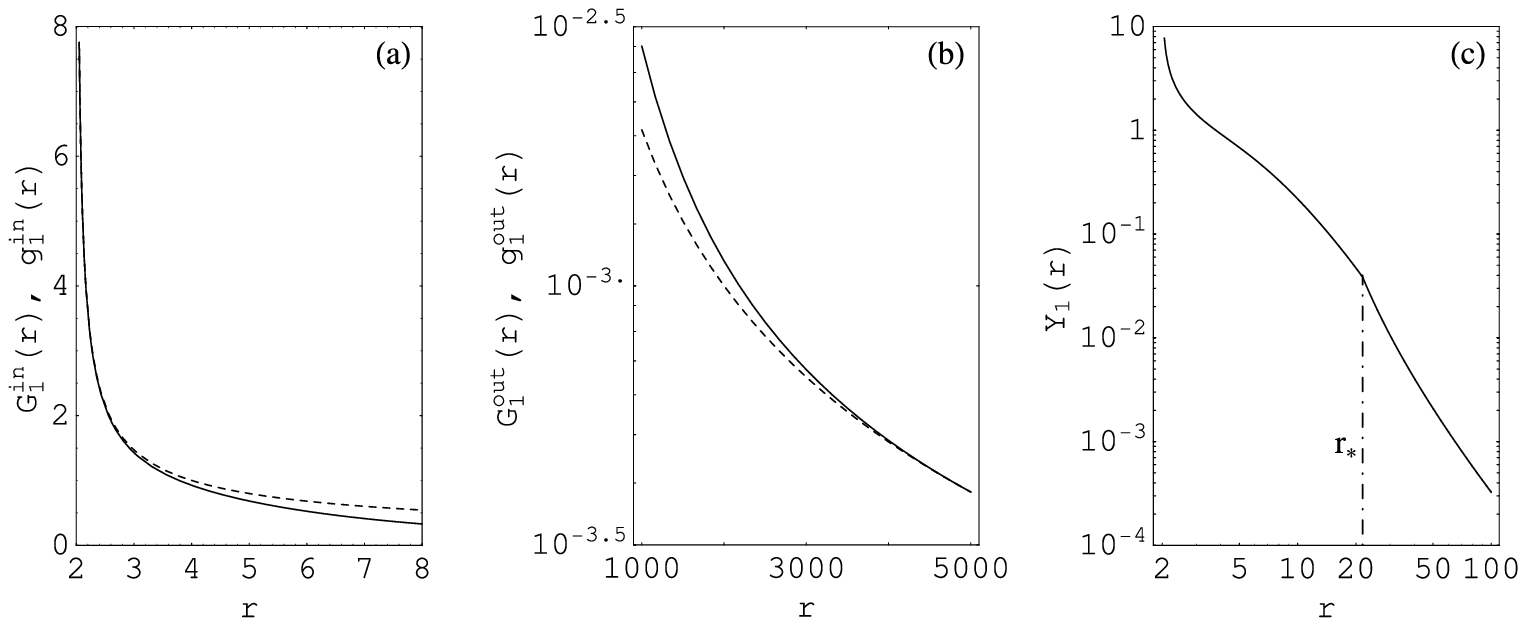}
\vskip-2.5cm
\caption{The fundamental solutions to eq.~(\ref{eq3.16}) $G_1^{\rm
in}(r)$ and $G_1^{\rm out}(r)$ ({\it solid lines}) for model~2 (M87) are
compared with the corresponding asymptotic solutions $g_1^{\rm in}(r)$
and $g_1^{\rm out}(r)$ ({\it dashed lines}) in panels {\it a} and {\it
b}, respectively (see eqs.~[\ref{eq3.23}]). The associated global
solution for the first eigenfunction $Y_1(r)$ (eq.~[\ref{eq3.22}]) is
plotted in panel {\it c}. The shock location at $r=r_*$ is indicated.}
\label{amod2}
\finfig
%



\hskip-1.5truein
\begin{deluxetable}{lccccccccr}
\tabletypesize{\scriptsize}
\tablecaption{Disk Structure Parameters\label{tbl-1}}
\tablewidth{0pt}
\tablehead{
\colhead{Source}
&\colhead{Model}
&\colhead{$M \atop (M_\odot)$}
&\colhead{$\dot M \atop (M_\odot \rm \ yr^{-1})$}
&\colhead{$\Ljet \atop \rm (ergs \ s^{-1})$}
& \colhead{$\epsilon_-$}
& \colhead{$\ell$}
& \colhead{$\epsilon_+$}
& \colhead{$r_*$}
& \colhead{$\mbox{\it \scriptsize u}_* \over \mbox{\it \scriptsize c}$}
}
\startdata
M87
& 2
& $3.0 \times 10^9$
& $1.3 \times 10^{-1}$
& $5.5 \times 10^{43}$
& $1.527 \times 10^{-3}$
& 3.1340
& $-5.746 \times 10^{-3}$
& 21.654
& 0.108
\\
\SgrA
& 5
& $2.6 \times 10^6$
& $8.8 \times 10^{-7}$
& $5.0 \times 10^{38}$
& $1.229 \times 10^{-3}$
& 3.1524
& $-8.749 \times 10^{-3}$
& 15.583
& 0.124
\\
\enddata


\hskip-1.0truein Note. -- All quantities are expressed in gravitational
units ($GM=c=1$) except as indicated.

\end{deluxetable}

\subsection{Numerical Solutions for the Eigenfunctions}

\begfig[t] \vskip-0.05cm \hskip-2.0cm
\includegraphics[width=7.0in]{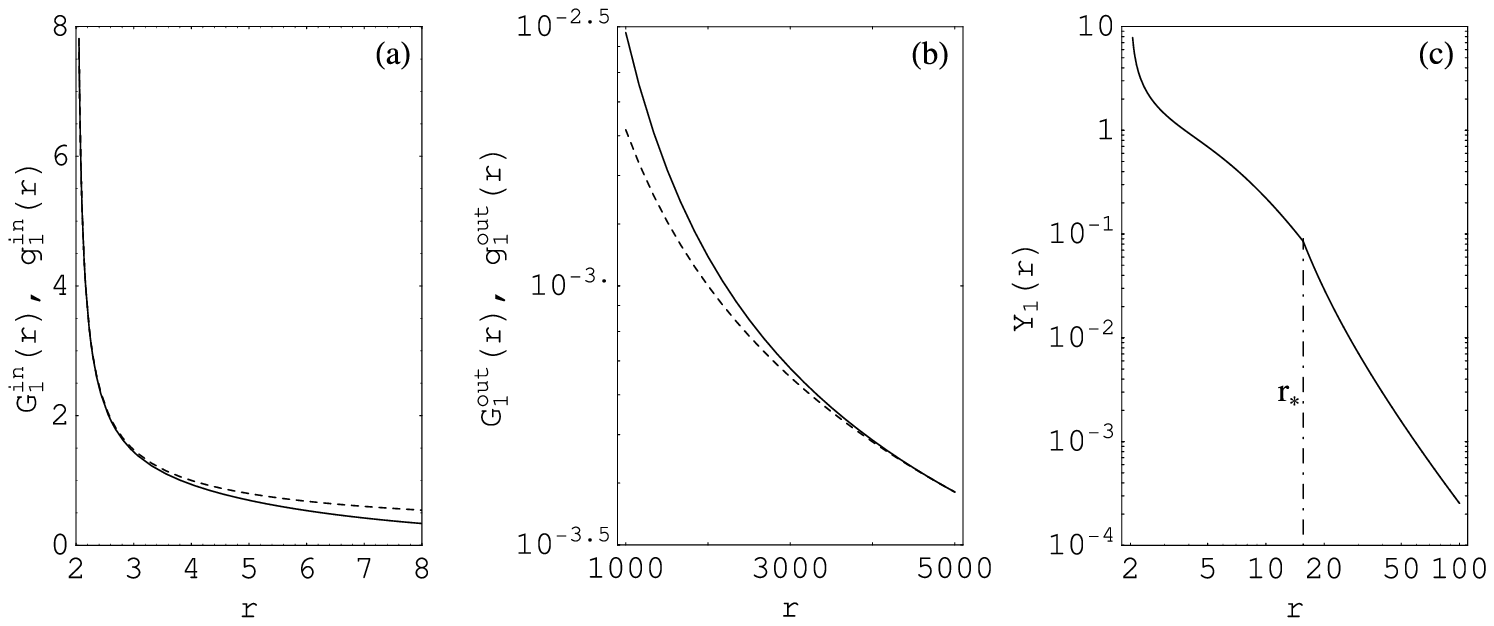}
\vskip-2.5cm
\caption{Same as Figure~\ref{amod2} but for model~5, which describes \SgrA.}
\label{amod5}
\finfig
\begfig[t] \centering
\vskip-0.0cm\hskip-0.4in
\plottwo{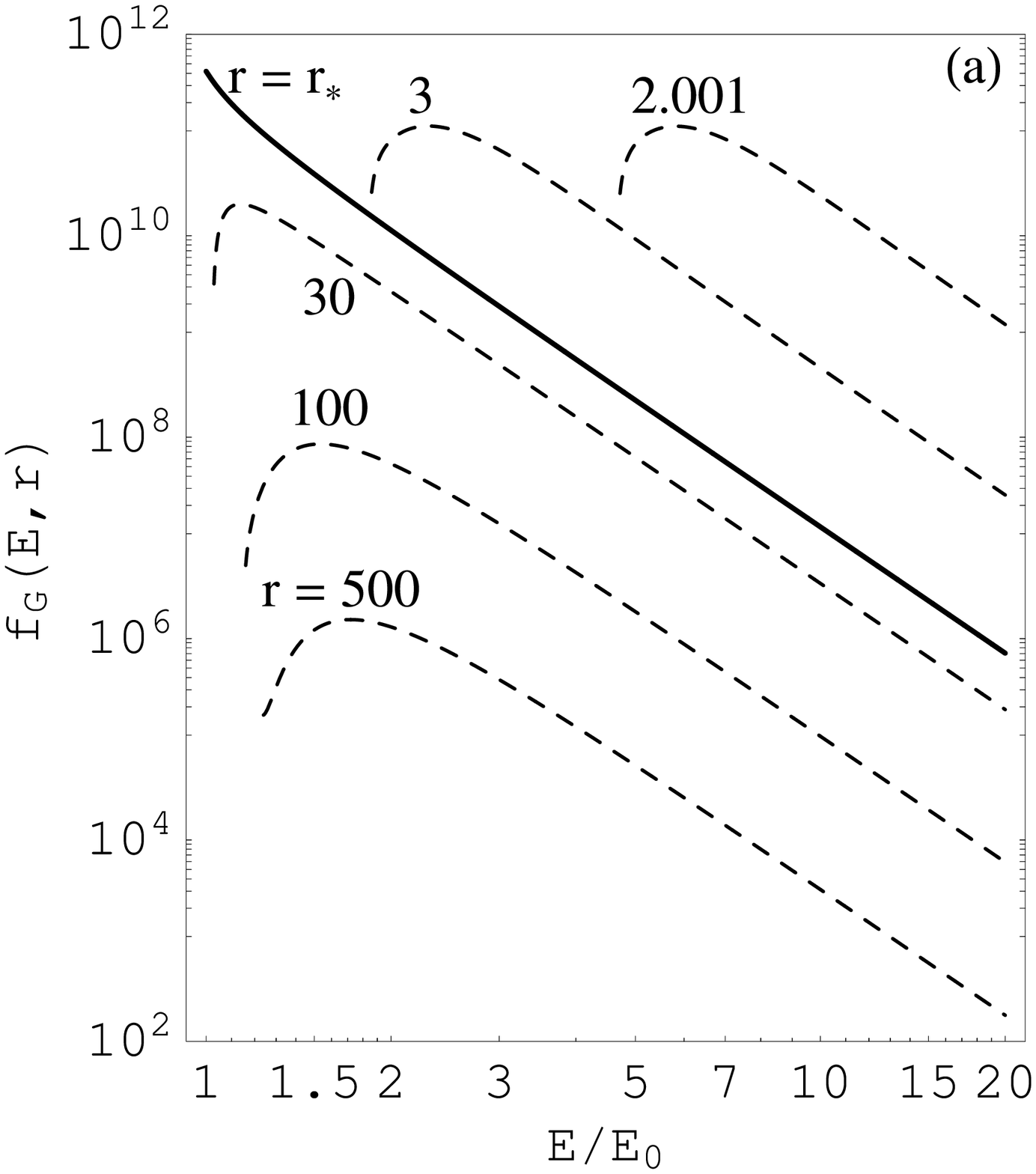}{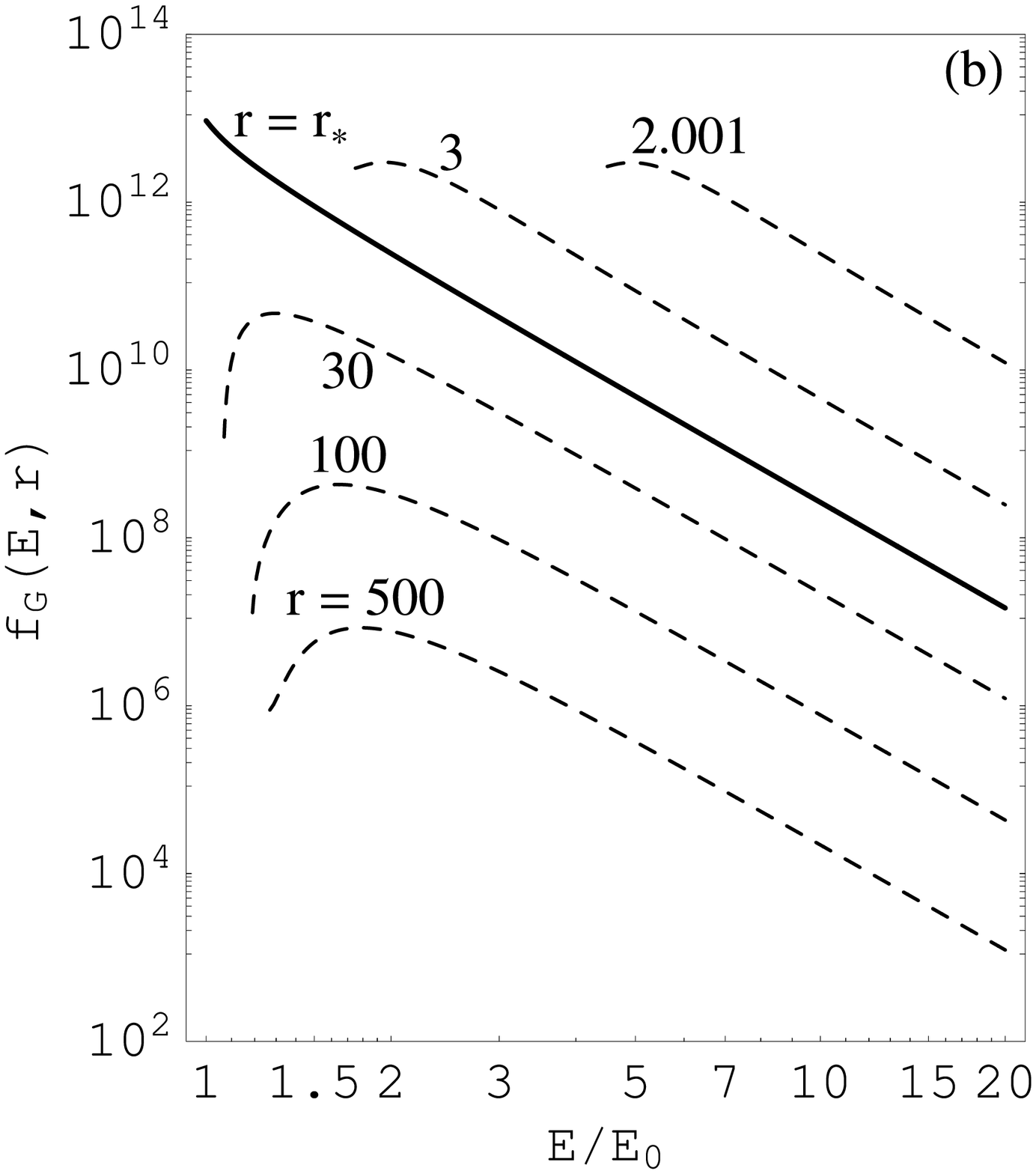} \vskip-1.cm
\caption{Results for the relativistic particle Green's function
$\green(E,r)$ in $\rm ergs^{-3} \, cm^{-3}$ computed using
eq.~(\ref{eq3.30}) for ({\it a}) M87 (model~2), and ({\it b}) \SgrA
(model~5). The value of the radius $r$ in units of $GM/c^2$ is indicated
for each curve, with $r=r_*$ denoting the shock location.}
\label{gfplot}
\finfig
\begfig[t]
\hskip-0.8cm \vskip-.0cm
\plottwo{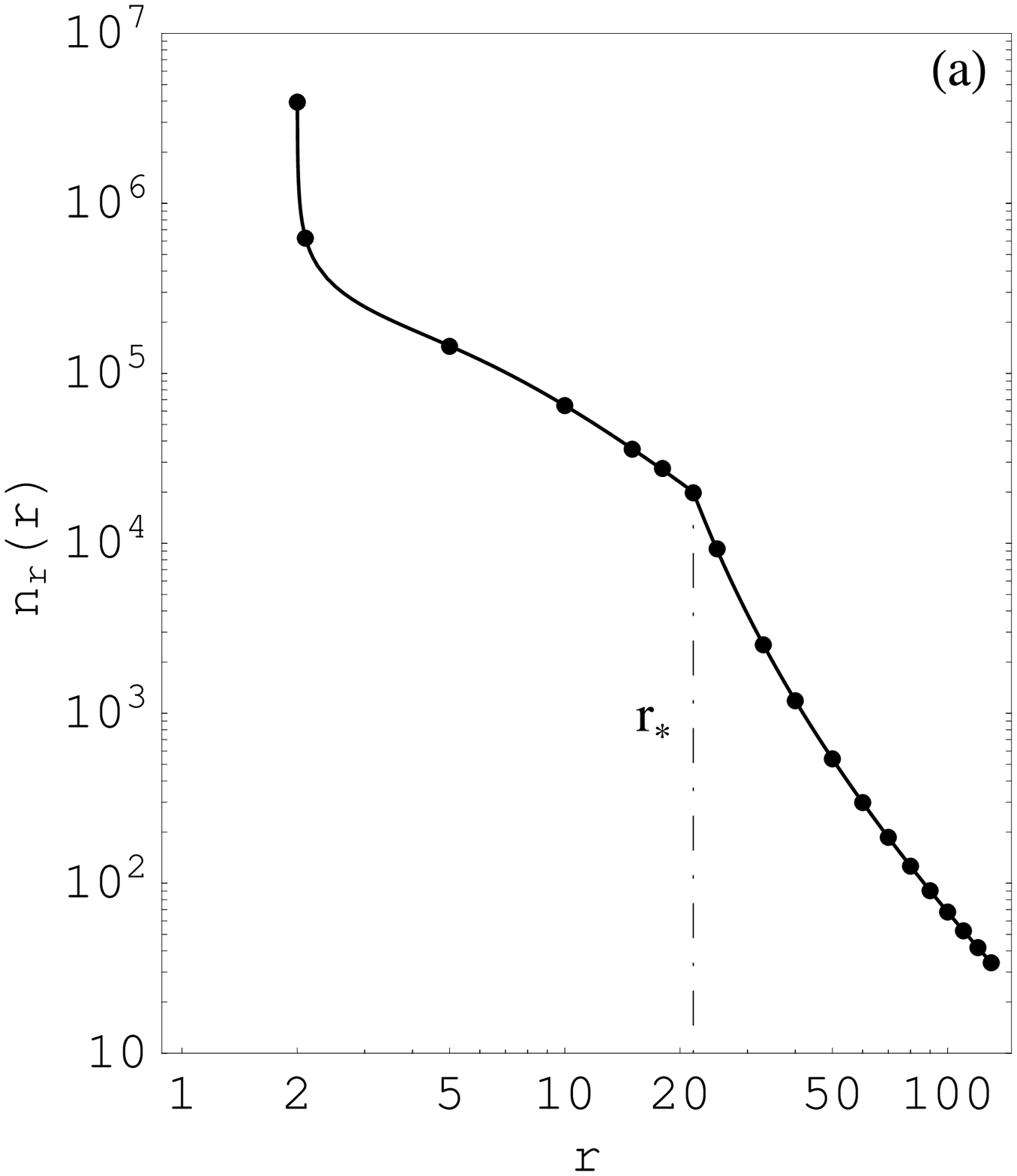}{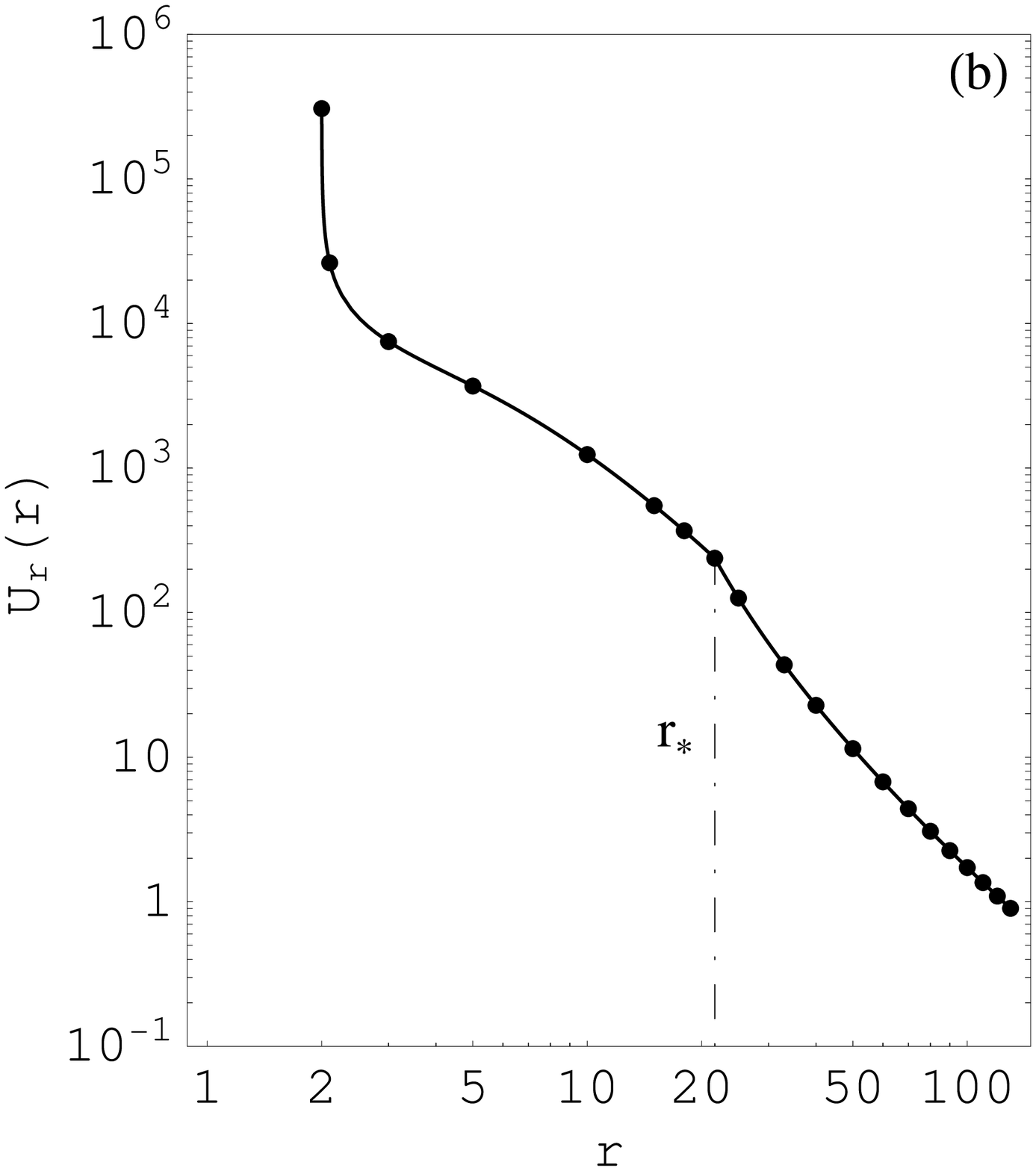}
\vskip-1.cm
\caption{Plots of ({\it a}) the relativistic particle number density and
({\it b}) the relativistic particle energy density in cgs units for
model~2, which describes M87. The solid lines represent the solutions
computed by numerically integrating the associated differential
equations, and the filled circles represent the corresponding results
obtained by integrating the Green's function using eqs.~(\ref{eq4.1}).
The location of the shock at $r=r_*$ is indicated. Note the close
agreement between the results, which confirms the accuracy of our
computational model.}
\label{nuplotmod2}
\finfig

In order to compute the Green's function $\green(E,r)$, we must first
solve for the eigenvalues $\lambda_n$ and eigenfunctions $Y_n(r)$
following the procedures described in \S\S~3.3 and 3.4 based on
equation~(\ref{eq3.18}). Once the eigenvalues and eigenfunctions have
been determined, we can compute the expansion coefficients $b_n$ using
equation~(\ref{eq3.37}), after which $\green(E,r)$ is evaluated using
equation~(\ref{eq3.30}). In Figure~\ref{amod2} we compare the
fundamental asymptotic solutions $G_n^{\rm in}(r)$ and $G_n^{\rm
out}(r)$ with the asymptotic solutions $g_n^{\rm in}(r)$ and $g_n^{\rm
out}(r)$ for model~2 with $n=1$ (see eqs.~[\ref{eq3.23}]). The
corresponding results for model~5 are plotted in Figure~\ref{amod5}. The
excellent agreement between the $G(r)$ and $g(r)$ functions confirms the
validity of the asymptotic relations we have employed near the event
horizon and also at large radii. Figures~\ref{amod2} and \ref{amod5}
also include plots of the global solutions for the first eigenfunction
$Y_1(r)$ associated with models~2 and 5, respectively (see
eq.~[\ref{eq3.22}]). Note that the global solutions display a derivative
jump at the shock location, as required according to
equation~(\ref{eq3.21}).

\subsection{Green's Function Particle Distribution}

We can combine our results for the eigenvalues, eigenfunctions, and
expansion coefficients to evaluate the Green's function $\green(E,r)$
using equation~(\ref{eq3.30}). In Figures~\ref{gfplot}a and
\ref{gfplot}b we plot $\green(E,r)$ as a function of the particle energy
$E$ at various radii $r$ in the disk for models~2 and 5, respectively.
Note that at the injection energy ($E=E_0$), the Green's function is
equal to zero except at the source/shock radius, $r=r_*$, in agreement
with equation~(\ref{eq3.36}). This reflects the fact that the particles
are rapidly accelerated after they are injected, and therefore all
particles have energy $E > E_0$ once they propagate away from the shock
location. The positive slopes observed at low energies when $r \ne r_*$
result from the summation of terms with either positive or
negative expansion coefficients $b_n$ (see \S~3.5). At energies above
the turnover, the spectrum has a power-law shape determined by the first
eigenvalue, $\lambda_1 \sim 4$ (see Fig.~2). The relatively flat slope
of the Green's function above the turnover reflects the high efficiency
of the particle acceleration in the shocked disk. By contrast, in
situations involving weak acceleration, the Green's function has a
strong peak at the injection energy surrounded by steep wings (e.g.,
Titarchuk \& Zannias 1998).

In Paper~2 we established that most of the injected particles are
advected inward by the accreting MHD waves, eventually crossing the
event horizon into the black hole. Hence only a small fraction of the
particles are able to diffuse upstream to larger radii, which is
confirmed by the strong attenuation of the particle spectrum with
increasing $r$ apparent in Figure~\ref{gfplot}. The particles in the
inner region ($r < r_*$) exhibit the greatest overall energy gain
because they are able to cross the shock multiple times, and they also
experience the strong compression of the flow near the event horizon.
This is consistent with the results for the mean energy distribution,
$\langle E \rangle \equiv U_r/n_r$, plotted in Figure~9 of Paper~2.

\subsection{Number and Energy Density Distributions}

In our numerical examples, we have generated results for the Green's
function $\green(E,r)$ by summing the first ten eigenfunctions using
equation~(\ref{eq3.30}). Once the Green's function energy distribution
has been determined at a given radius, we can integrate it with respect
to the particle energy $E$ to obtain the corresponding values for the
number and energy densities. By combining equations~(\ref{eq3.7}) and
(\ref{eq3.30}) and integrating term-by-term, we obtain
\begeq
n^{\rm G}_r(r) \equiv 4 \pi E_0^3 \sum^{N_{\rm max}}_{n=1}
{b_n Y_n(r) \over \lambda_n-3} \ ,
\ \ \ \ \
U^{\rm G}_r(r) \equiv 4 \pi E_0^4 \sum^{N_{\rm max}}_{n=1}
{b_n Y_n(r) \over \lambda_n-4} \ .
\label{eq4.1}
\fineq
Equations~(\ref{eq4.1}) provide an important tool for checking the
self-consistency of our entire formalism. This is accomplished by
comparing the results for the number and energy densities computed using
equations~(\ref{eq4.1}) with the corresponding values obtained by
directly solving the differential equations for $n_r$ and $U_r$ using
the method developed in Paper~2. In principle, the two sets of results
should agree closely if our procedure for computing $\green(E,r)$ is
robust. We compare the model~2 results obtained for $n_r$ and $U_r$ in
Paper~2 with those computed using equations~(\ref{eq4.1}) in
Figure~\ref{nuplotmod2}. The corresponding comparison for model~5 is
carried out in Figure~\ref{nuplotmod5}. Note the close agreement between
the two sets of results, which confirms the validity of the analysis
involved in searching for the eigenvalues, computing the eigenfunctions,
solving for the expansion coefficients, and calculating the Green's
function. Moreover, the results in Figures~6 and 7 indicate that
equation (\ref{eq3.30}) converges successfully using the first ten terms
($N_{\rm max}=10$) in the series.

\begfig[t]
\hskip-0.8cm \vskip-.0cm
\plottwo{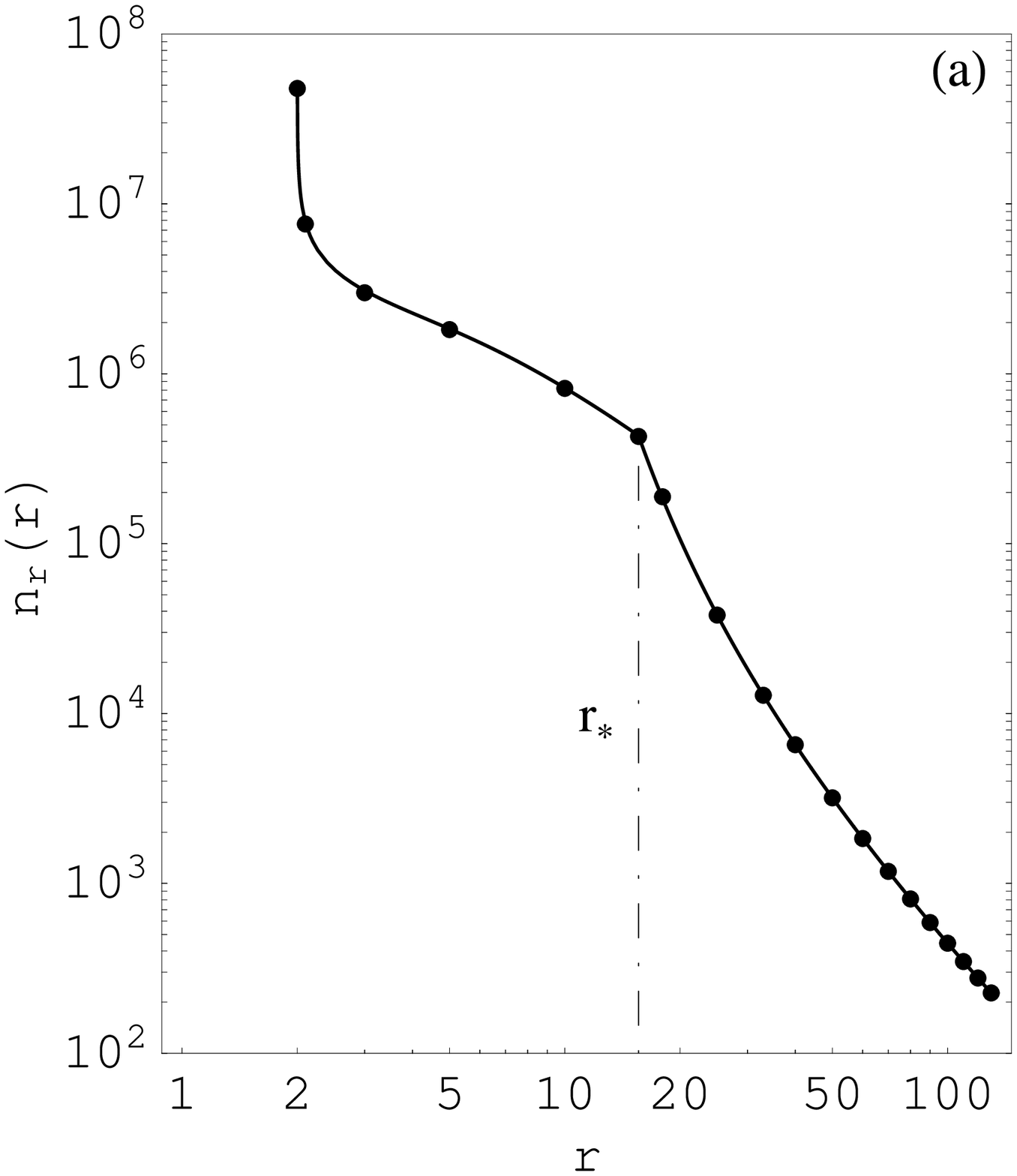}{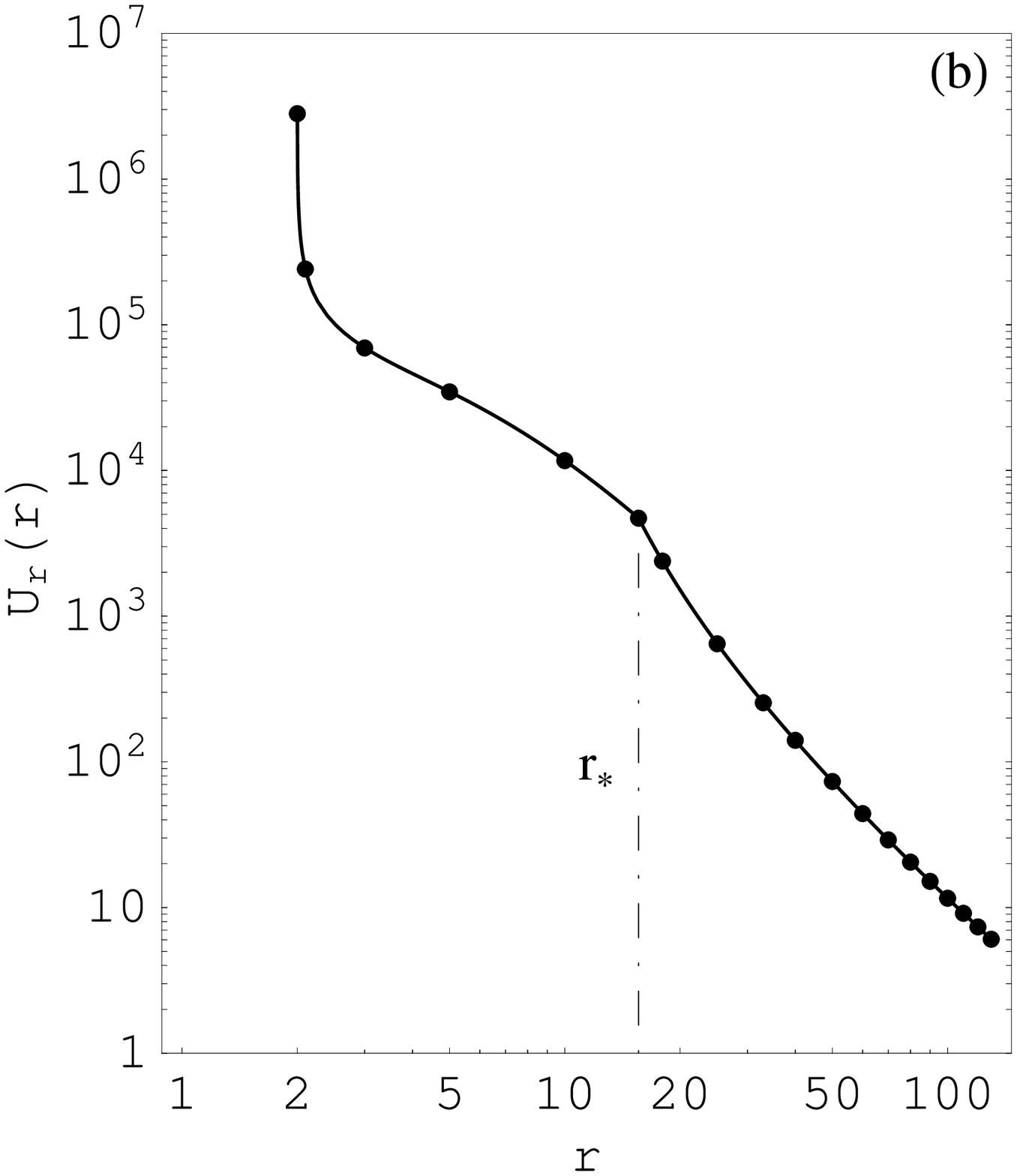}
\vskip-1.cm
\caption{Same as Figure~\ref{nuplotmod2} but for model~5, which describes
\SgrA.}
\label{nuplotmod5}
\finfig

\subsection{Escaping Particle Distribution}

The Green's function $\green(E,r)$ computed using
equation~(\ref{eq3.30}) represents the energy distribution of the
relativistic particles inside the disk at radius $r$. However, our
primary goal in this paper is the calculation of the energy distribution
of the relativistic particles {\it escaping from the disk} at the shock
radius, $r_*$. In order to compute the energy spectrum of the escaping
particles, we need to integrate equation~(\ref{eq3.4}) over the volume
of the disk, which yields
\begeq
\dot N_E^{\rm esc} = (4 \pi E)^2 r_* H_* \, c \, A_0
\green(E,r_*) \ ,
\label{eq4.2}
\fineq
where $\dot N_E^{\rm esc} \, dE$ represents the number of particles
escaping from the disk per unit time with energy between $E$ and $E+dE$.
At the highest particle energies, the escaping particle number spectrum
has a power-law shape, with $\dot N_E^{\rm esc} \propto E^{-\alpha}$,
where $\alpha = \lambda_1 -2$ and $\lambda_1$ is the first eigenvalue
(see eqs.~[\ref{eq3.30}] and [\ref{eq4.2}]). The total number of
particles escaping from the disk per second, $\dot N_{\rm esc}$, is
related to the spectrum $\dot N_E^{\rm esc}$ via
\begeq
\dot N_{\rm esc} = \int^\infty_0 \dot N_E^{\rm esc} dE
= 4 \pi r_* H_* \, c \, A_0 \, n_* \ ,
\label{eq4.3}
\fineq
where $n_* \equiv n_r(r_*)$ is the relativistic particle number density
at the shock location. The corresponding result for the total energy
escape rate is given by
\begeq
\Lesc \equiv \int^\infty_0 \dot N_E^{\rm esc} \, E \, dE
= 4 \pi r_* H_* \, c \, A_0 \, U_* \ ,
\label{eq4.4}
\fineq
where $U_* \equiv U_r(r_*)$ is the relativistic particle energy density
at the shock location.

The escaping particle distributions for M87 (model~2) and \SgrA
(model~5) are plotted in Figure~\ref{gfesc}. The results obtained in the
case of M87 are about five orders of magnitude larger than those
associated with \SgrA~ due to the corresponding difference in the
reported jet luminosity $\Ljet$ for these two sources (see
Table~\ref{tbl-1}). In Table~\ref{tbl-2} we list the values obtained for
$\dot N_0$, $\dot N_{\rm esc}$, $n_*$, $U_*$, $\kappa_*$, $\kappa_0$,
$A_0$, $\Eesc$, and $\Gamma_\infty$ in models~2 and 5, where $\Eesc
\equiv U_*/n_*$ denotes the mean energy of the particles escaping from
the disk to power the jet, and $\Gamma_\infty$ represents the terminal
Lorenz factor of the jet. We note that the values obtained for $\Lesc$
using the integral in equation~(\ref{eq4.4}) are in excellent agreement
with the values for $\Ljet$ reported in Paper~2 and listed in
Table~\ref{tbl-1}. The agreement between $\Lesc$ and $\Ljet$ confirms
that our model satisfies global energy conservation. We also find that
the values for $\dot N_{\rm esc}$ obtained using the integral in
equation~(\ref{eq4.3}) are in excellent agreement with the values listed
in Table~\ref{tbl-2}, which were taken from Paper~2. These tests confirm
the validity of the analytical approach we have utilized in our
derivation of the Green's function, given by equation~(\ref{eq3.30}).



\hskip-1.5truein
\begin{deluxetable}{cccccccccr}
\tabletypesize{\scriptsize}
\tablecaption{Transport Equation Parameters\label{tbl-2}}
\tablewidth{0pt}
\tablehead{
\colhead{Model}
& \colhead{$\dot N_0 \atop \rm (s^{-1})$}
& \colhead{$\dot N_{\rm esc} \over \dot N_0$}
& \colhead{$\mbox{\it \scriptsize n}_* \atop \rm(cm^{-3})$}
& \colhead{$\mbox{\it \scriptsize U}_* \atop \rm(ergs \, cm^{-3})$}
& \colhead{$\kappa_*$}
& \colhead{$\kappa_0$}
& \colhead{$A_0$}
& \colhead{$E_{\rm esc} \over E_0$}
& \colhead{$\Gamma_\infty$}
}
\startdata
2
& $2.75 \times 10^{46}$
& 0.17
& $2.01 \times 10^4$
& $2.39 \times 10^2$
& 0.427877
& 0.02044
& 0.0124
& 5.95
& 7.92
\\
5
& $2.51 \times 10^{41}$
& 0.18
& $4.33 \times 10^5$
& $4.71 \times 10^3$
& 0.321414
& 0.02819
& 0.0158
& 5.45
& 7.26
\\
\enddata


\hskip-1.0truein Note. -- All quantities are expressed in gravitational
units ($GM=c=1$) except as indicated.

\end{deluxetable}

\begfig[t]
\hskip-0.8cm \vskip-.0cm
\plottwo{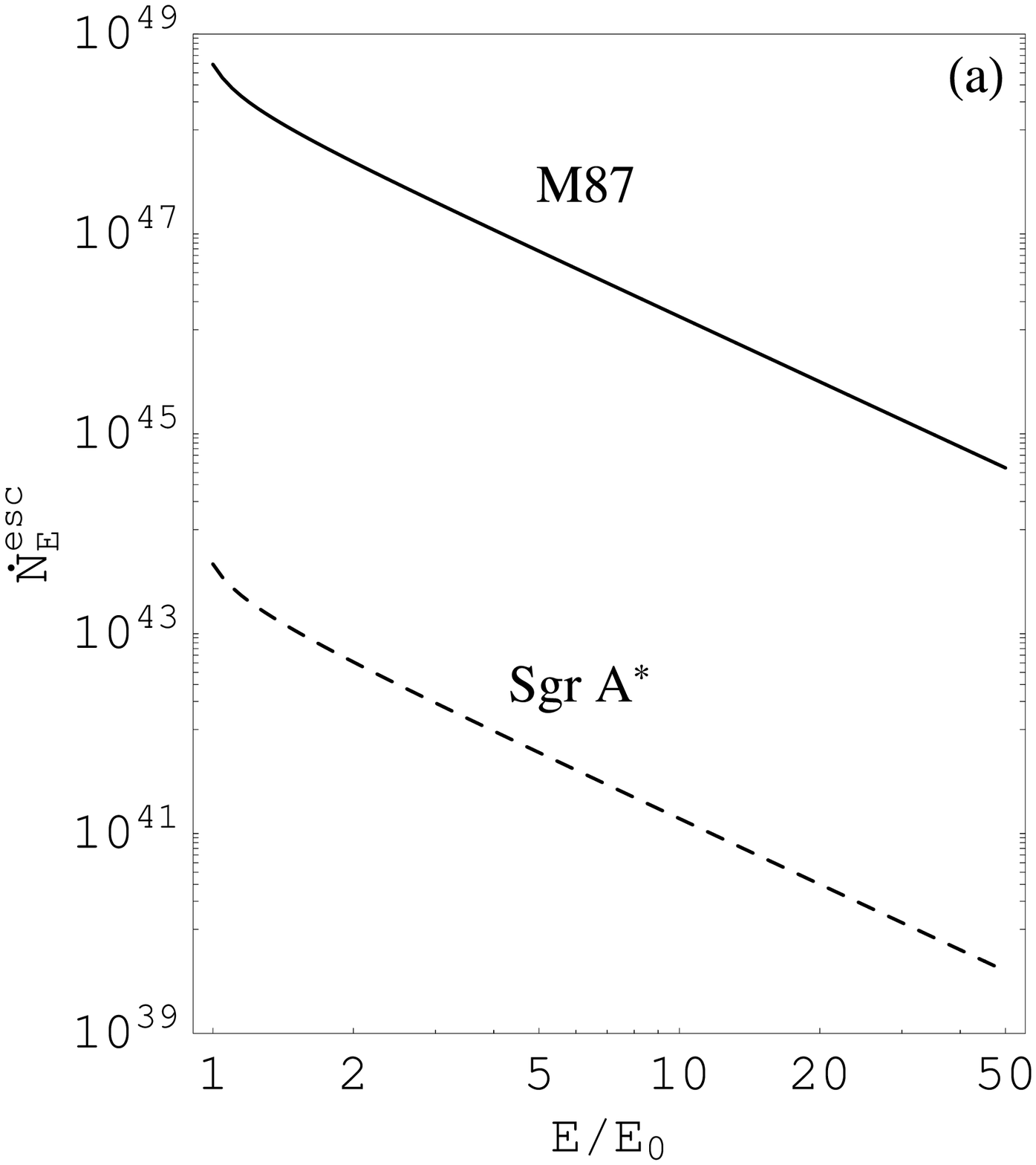}{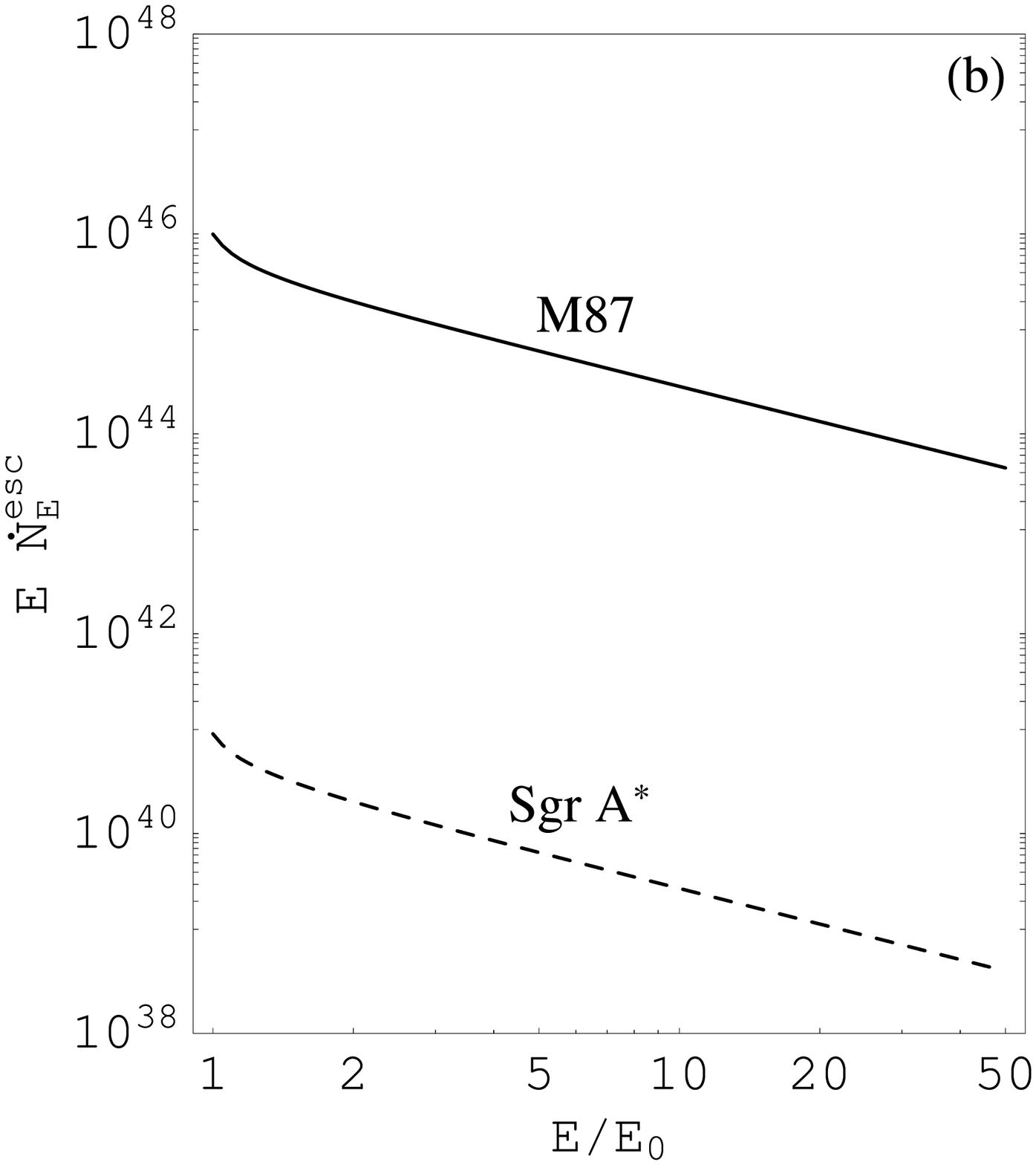}
\vskip-1.cm
\caption{Plots of ({\it a}) the number distribution $\dot N_E^{\rm esc}$
and ({\it b}) the energy distribution $E \, \dot N_E^{\rm esc}$ for the
relativistic particles escaping at the shock location, $r=r_*$ (see
eq.~[\ref{eq4.2}]). The solid curves represent model~2 (M87), and the
dash curves represent model~5 (\SgrA).}
\label{gfesc}
\finfig

\section{OBSERVATIONAL IMPLICATIONS}

The jets of relativistic particles produced by the first-order Fermi
shock acceleration mechanism considered here will flow outwards from the
disk with initially high pressure, and low velocity. The rapid expansion
of the plasma will lead to strong acceleration as the gas cools and the
random internal energy is converted into ordered kinetic energy (see
Paper~2). The terminal Lorenz factor, $\Gamma_\infty$, associated with
the M87 (model~2) and \SgrA (model~5) jets considered here are found to
be 7.92 and 7.26, respectively, based on comparison with the observed
jet kinetic luminosities. However, robust confirmation of the model
requires a detailed comparison between the predicted emission and the
observed multifrequency spectra for a variety of sources. The jets
envisioned here are expected to be charge neutral, with the protons
``dragging'' along an equal number of electrons. The electrons are
expected to cool rapidly via inverse-Compton and bremsstrahlung emission
produced while the particles are still close to the active nucleus. On
the other hand, the relativistic protons will cool inefficiently via
direct radiation, and they also lose very little energy to the electrons
via Coulomb coupling due to the low gas density. Hence it is expected
that the ions will maintain their high energies until they collide with
particles or radiation in the ambient medium. Detailed consideration of
this issue is beyond the scope of the present paper, and will be pursued
in future work. However, we briefly discuss a few of the probable
scenarios for the production of the observed radio, X-ray, and
$\gamma$-ray emission by the jet particles below.

{\it Scenario 1:} Anyakoha et al. (1987) and Morrison et al. (1984) have
suggested that high-energy protons in the jets may undergo proton-proton
collisions with the ambient medium, resulting in the production of
charged and neutral pions. The neutral pions subsequently decay into
$\gamma$-rays, and the charged pions decay into electron-positron pairs
which produce further $\gamma$-rays via inverse-Compton and annihilation
radiation. These processes can explain the observed radio and
$\gamma$-ray emission in blazar sources such as 3C~273. Anyakoha et al.
(1987) demonstrated that the production of the observed $\gamma$-rays
requires protons with Lorentz factors in the range $1.4 \lapprox \gamma
\lapprox 86$, and the production of the observed radio emission requires
protons with Lorentz factors in the range $7 \lapprox \gamma \lapprox
300$. Based on our model, we find that the terminal proton Lorentz
factor, $\Gamma_\infty$, for the jets in M87 and \SgrA~ is given by
$\Gamma_\infty \sim 8$ and $\Gamma_\infty\sim 7$, respectively (see
Table~\ref{tbl-2}). We therefore conclude that the escaping protons
accelerated in our model reach energies high enough to produce the
observed radio and $\gamma$-ray emission in M87 and \SgrA. In addition
to the radiation emitted by the escaping, outflowing protons, the
relativistic protons remaining inside the disk will produce additional
$\gamma$-rays (e.g., Eilek \& Kafatos 1983; Bhattacharyya et al. 2003,
2006), although the escape of these photons from the disk may be
problematic due to $\gamma$-$\gamma$ attenuation (Becker \& Kafatos
1995).

{\it Scenario 2:} Usov \& Smolsky (1998) suggested that the nonthermal
$\gamma$-ray and X-ray emission from AGN jets may be produced by
energetic electrons accelerated due to collisions between
relativistically moving jet clouds and the ambient medium. The
collisions accelerate the electrons up to energies $\sim
m_p\,c^2\,\Gamma_\infty$, where $\Gamma_\infty$ is the terminal Lorentz
factor of the jet in our model. Gamma-rays are produced when the
high-energy electrons scatter external radiation. Based on the results
presented by Usov \& Smolsky (1998), we find that the mean energy of the
scattered photons emerging from the jet in this scenario is
$\left<\varepsilon_\gamma\right> \lapprox 400 \, \Gamma^2_\infty\,$keV,
where $\Gamma_\infty \sim 7-8$ (see Table~\ref{tbl-2}). Additional
X-rays and $\gamma$-rays may also be produced as a result of direct
upscattering of background photons by beamed electrons if the Lorentz
factor $\Gamma_\infty \sim 10$, which is comparable to the range
predicted by our model (e.g., Dermer \& Schlickeiser 1993).

{\it Scenario 3:} This last scenario is quite different from the first
two. Sikora \& Madejski (2000) argue that the jet in optically
violently variable (OVV) quasars contains a mixture of compositions,
since the pure $e^+$ $e^-$ jets overproduce soft X-rays and the pure
$e-p$ jets underproduce nonthermal X-rays. To resolve these issues, they
suggest that initially $e-p$ plasma is injected at the base of the jets,
with the protons providing the inertia to account for the kinetic
luminosity of the jet. This is consistent with our model of
proton/electron escape at the shock. Subsequently, $e^+$ $e^-$ pairs are
produced via interactions with the hard X-rays/soft $\gamma$-rays
emitted from the hot accretion disk corona. This leads to velocity and
density perturbations in the jet and also to the formation of shocks,
where the particles are further accelerated. The consequence of the
Sikora \& Madejski (2000) model is that pair production in the
accretion-disk corona can be responsible for the fast ($\sim {\rm day}$)
variability observed in OVV quasars at very small angles relative to the
jet axis, and also detected in the MeV-GeV $\gamma$-ray regime. We can
expect a similar effect in our theoretical model since the escape of the
relativistic protons/electrons also occurs at the base of the jet, at
the shock location. Our model therefore provides a natural connection
between the particle acceleration mechanism operating in the disk and
the formation of the energetic outflows analyzed by Sikora \& Madejski
(2000).

\section{CONCLUSIONS}

In this paper we have demonstrated that particle acceleration at a
standing, isothermal shock in an ADAF disk can energize the relativistic
protons that power the jets emanating from radio-loud sources containing
black holes. The work presented here represents a new type of synthesis
that combines the standard model for a transonic ADAF flow with a
self-consistent treatment of the relativistic particle transport
occurring in the disk. The energy lost from the background (thermal) gas
at the isothermal shock location results in the acceleration of a small
fraction of the background particles to relativistic energies. One of
the major advantages of our coupled, global model is that it provides a
single, coherent explanation for the disk structure and the formation of
the outflows based on the well-understood concept of first-order Fermi
acceleration at shock waves. The theory employs an exact mathematical
approach in order to solve simultaneously the coupled hydrodynamical and
particle transport equations.

The shock acceleration mechanism analyzed in this paper is effective
only in rather tenuous, hot disks, and therefore we conclude that our
model may help to explain the observational fact that the brightest X-ray
AGNs do not possess strong outflows, whereas the sources with low X-ray
luminosities but high levels of radio emission do. We suggest that the
gas in the luminous X-ray sources is too dense to allow efficient Fermi
acceleration of a relativistic particle population, and therefore in
these systems, the gas simply heats as it crosses the shock. Conversely,
in the tenuous ADAF accretion flows studied here, the relativistic
particles are able to avoid thermalization due to the long collisional
mean free path, resulting in the development of a significant nonthermal
component in the particle distribution which powers the jets and
produces the strong radio emission.

The transport model we have utilized includes the effects of spatial
diffusion, first-order Fermi acceleration, bulk advection, and vertical
escape through the upper and lower surfaces of the disk. All of the
theoretical parameters can be tied down for a source with a given mass,
accretion rate, and jet luminosity. The diffusion coefficient model
employed in this work is consistent with the behavior expected as
particles approach the event horizon and also at large radii. Close to
the event horizon, inward advection at the speed of light dominates over
outward diffusion, as expected, while at large $r$, diffusion dominates
over advection. The vertical escape of the energetic particles through
the upper and lower surfaces of the disk at the shock location occurs
via spatial diffusion in the tangled magnetic field. The approach taken
here closely parallels the early studies of cosmic-ray shock
acceleration. As in those first investigations (e.g., Blandford \&
Ostriker 1978), we have employed the test particle approximation in
which the pressure of the accelerated particles is assumed to be
negligible compared with that of the thermal background gas. This
approximation was shown to be valid for our specific applications to M87
and \SgrA in Paper~2.

We have computed the Green's function describing the particle
distribution inside the disk, as well as the energy distribution for the
relativistic particles that escape at the shock location to power the
jets. The escaping particle distributions for model~2 (M87) and model~5
(\SgrA) each possess high-energy power-law tails, as expected for a
Fermi mechanism. The relatively flat slope of the energy distribution
also implies the potential for an interesting connection between the
acceleration process explored here and the formation of the universal
cosmic-ray background. This is consistent with the results of Szabo \&
Protheroe (1994), who demonstrated that AGN may be an important source
of cosmic rays in the region of the ``knee.'' The coupled,
self-consistent model for the disk structure, the acceleration of the
relativistic particles, and the production of the outflows presented in
this paper represents the first step towards a fully integrated
understanding of the relationship between accretion dynamics and the
formation of the relativistic jets commonly observed around radio-loud
compact sources.

The authors would like to thank the referee, Lev Titarchuk, for
providing several insightful comments that led to significant
improvements in the manuscript.

\newpage

\section*{APPENDICES}

\appendix

\section{Orthogonality of the Spatial Eigenfunctions}

We can establish the orthogonality of the spatial eigenfunctions
$Y_n(r)$ by starting with the Sturm-Liouville form (see
eq.~[\ref{eq3.25}])
\begeq
{d \over dr}\left[S(r) \, {dY_n \over dr} \right]
+ \lambda_n \, \omega(r) \, Y_n(r) = 0 \ ,
\label{Aeq1}
\fineq
where $\omega(r)$ and $S(r)$ are given by equations~(\ref{eq3.27}) and
(\ref{eq3.26}), respectively. Let us suppose that $\lambda_n$ and
$\lambda_m$ denote two distinct eigenvalues ($\lambda_n \ne \lambda_m$)
with associated spatial eigenfunctions $Y_n(r)$ and $Y_m(r)$. Since
$Y_n$ and $Y_m$ each satisfy equation~(\ref{Aeq1}) for their respective
values of $\lambda$, we can write
\begeqarray
Y_n(r) \left\{{d \over dr}\left[S(r) \, {dY_m \over dr}
\right] + \lambda_m \, \omega(r) \, Y_m(r) \right\}
& = & 0
\label{Aeq2} \\
Y_m(r) \left\{{d \over dr}\left[S(r) \, {dY_n \over dr}
\right] + \lambda_n \, \omega(r) \, Y_n(r) \right\} & = & 0 \, .
\label{Aeq3}
\fineqarray
Substracting equation~(\ref{Aeq3}) from equation~(\ref{Aeq2}) yields
\begeq
Y_n(r) \, {d \over dr}\left[S(r) \, {dY_m \over dr}\right]
- Y_m(r) \, {d \over dr}\left[S(r) \, {dY_n \over dr} \right]
= (\lambda_n-\lambda_m) \, \omega(r) \, Y_n(r) \, Y_m(r) \ .
\label{Aeq4}
\fineq
We can integrate equation~(\ref{Aeq4}) by parts from $r = \rs$ to
$r = \infty$ to obtain
\begeqarray
Y_n(r) \, S(r) \, {dY_m \over dr} \bigg|^\infty_{\rs}
- \int^\infty_{\rs} S(r) \, {dY_m \over dr}
\, {dY_n \over dr} \, dr \nonumber \\
- Y_m(r) \, S(r) \, {dY_n \over dr} \bigg|^\infty_{\rs}
+ \int^\infty_{\rs} S(r) \, {dY_n \over dr}
\, {dY_m \over dr} \, dr \nonumber \\
= (\lambda_n-\lambda_m) \int^\infty_{\rs} \omega(r) \, Y_n(r) \, Y_m(r)
\, dr \ .
\label{Aeq5}
\fineqarray
Upon cancellation of like terms, this expression reduces to
\begeq
S(r) \left[ Y_n(r) {dY_m \over dr} - Y_m(r)
{dY_n \over dr} \right]^\infty_{\rs}
= (\lambda_n-\lambda_m) \int^\infty_{\rs} \omega(r) \, Y_n(r) \, Y_m(r)
\, dr \ .
\label{Aeq6}
\fineq
Based upon the asymptotic behaviors of $Y_n(r)$ as $r \to \rs$ and
$r \to \infty$ given by equation~(\ref{eq3.23}), we conclude that
the left-hand side of equation~(\ref{Aeq6}) vanishes, leaving
\begeq
\int^\infty_{\rs} \omega(r) \, Y_m(r) \, Y_n(r) \, dr = 0 \ ,
\ \ \ \ \ \ \ m \neq n \ .
\label{Aeq7}
\fineq
This result establishes that $Y_m$ and $Y_n$ are {\it orthogonal
eigenfunctions} relative to the weight function $\omega(r)$ defined
in equation~(\ref{eq3.27}).

\end{document}